\documentclass[useAMS,usenatbib,usegraphicx]{mn2e}  
\usepackage[usenames]{color}

\newcommand{\be}{\begin{equation}}
\newcommand{\ee}{\end{equation}}
\newcommand{\bea}{\begin{eqnarray}}
\newcommand{\eea}{\end{eqnarray}}

\title[Dark matter in MACSJ0416.1-2403] 
{Free Form Lensing Implications for the Collision of Dark Matter and Gas 
in the Frontier Fields Cluster MACSJ0416.1-2403}  

\author[Diego et al.]  
  {Jose M. Diego$^1$\footnote{jdiego@ifca.unican.es}, 
   T. Broadhurst$^{2,3}$, 
   S. M. Molnar$^{4}$
   D. Lam$^{5}$,
   J. Lim$^{5}$, \\
$^{1}$IFCA, Instituto de F\'isica de Cantabria (UC-CSIC), Av. de Los Castros 
s/n, 39005 Santander, Spain\\
$^{2}$Fisika Teorikoa, Zientzia eta Teknologia Fakultatea, Euskal Herriko 
Unibertsitatea UPV/EHU\\ 
$^{3}$IKERBASQUE, Basque Foundation for Science, Alameda Urquijo, 36-5 48008 
Bilbao Spain\\
$^{4}$ Department of physics, National Taiwan University, Taipei 10617, Taiwan.\\
$^{5}$ Department of Physics, The University of Hong Kong, Pokfulam Road, Hong Kong\\
}

\date{Draft version \today}  
  
\pagerange{\pageref{firstpage}--\pageref{lastpage}}  
  
\begin{document}  
\maketitle  
 
\label{firstpage}  
\begin{abstract}  

We present a free form mass reconstruction of the massive lensing
cluster MACSJ0416.1-2403 using the latest Hubble Frontier Fields
data. Our model independent method finds that the extended lensing
pattern is generated by two elongated, closely projected clusters of
similar mass. Our lens model identifies new lensed images with which
we improve the accuracy of the dark matter distribution. We find that
the bimodal mass distribution is nearly coincident with the bimodal
X-ray emission, but with the two dark matter peaks lying closer
together than the centroids of the X-ray emisison. We show this can be
achieved if the collision has occurred close to the plane and such
that the cores are deflected around each other. The projected mass
profiles of both clusters are well constrained because of the many
interior lensed images, leading to surprisingly flat mass profiles of
both components in the region $15-100$kpc. We discuss the extent to
which this may be generated by tidal forces in our dynamical model
which are large during an encounter of this type as the cores {\it graze}
each other.  The relative velocity between the two cores is estimated 
to be about 1200 km/s and mostly along the line of sight 
so that our model is consistent with the relative redshift difference 
between the two cD galaxies ($\delta z \approx 0.04$).
\end{abstract}  

\section{Introduction}\label{sect_intro}  

 Strong gravitational lensing is entering a new era of enhanced
 precision with the arrival of the new Hubble data from the CLASH
 \cite{Postman2012b} and now from the Frontier
 Fields\footnote{http://www.stsci.edu/hst/campaigns/frontier-fields/}
 (HFF, hereafter) programs. 
 Clusters observed under these programs
 routinely contain tens of multiply lensed systems with well defined
 colours to help identify counter-images. 
 The cluster RXJ1532.8+3021 is a notable exception where CLASH data does not reveal 
 any strongly lensed system. Weak lensing analysis suggests a very low 
 concentration for this cluster \cite{Merten2014}. In X-rays, the cluster shows 
 two prominent cavities in opposite directions, a clear signature of a very 
 powerfull black hole in the central galaxy supported by VLA radio observations 
 of the associated jets. The picture in this cluster is consistent with the 
 existence of a supermassive blackhole with a mass $M_{BH} \sim 10^{10} 
 M_{\odot}$ \cite{Hlavaceck2013}.

 The abundance of strong
 lensing observables allows for an unprecedentedly detailed
 reconstruction of the dark matter distribution that is responsible
 for the lensing distortions. The lensing models have evolved to take
 increasing advantage of the improvmentin data quality. The most
 massive clusters, with the largest quantity of strong lensing
 information, are often the most complex, undergoing dymanical
 evolution. Paramettric models are most powerful for relaxed systems
 but arbitrary choices have to be made when representing substructures
 with idealized halo components 
\cite{Kneib1996,Sharon2005,Halkola2006,Limousin2007,Nakajima2007,Nieuwenhuizen2013}
 
We have tacked this problem in two ways previously, with application
to A1689. Firstly, by a compromise {\it semi-parametric} method which
assumed mass approximately traces light \cite{Broadhurst05a} has
been sucessful in dealing with this problem, and useful for located
many sets of multiple images and has been developed further for the
clash data \cite{Zitrin2011b,Zitrin2012,Zitrin2013}  but is still does not allow for the
possibility that the dynamics of the dark may differ from that of the
member galaxies. A fully general free-form approach was deveoped by
\cite{Diego05a,Diego05b,Diego2007}, based on a pixellated grid representing the lens
plane. This does not provide sufficient accuracy to find new systems
so that it must rely on lensed systems found by other models.
More recently we have found it very advantagous to incororate the
known position of member galaxies in the grid solution, so that
meaningful solutions can be found \cite{Sendra2014,Diego2014},
and here we emply this method to model the complex merging cluster
MACSJ0416.1-2403 (MACSJ0416 hereafter), relying on our own image identifications.
free form approach of dividding 

 The success of the above approaches has relied also on a better
 understanding of the optimization of colour information. By utilising
 overlapping broad bands of the CLASH program covering the UV to the
 NIR, we can maximize the photometric redshift accuracy possible with Hubble
 and have provided reliable examples of the most distant galaxies
 known, as in the case of the $z\simeq 11$ candidate lensed by
 MACS0647 \citep{Coe2013}, where multiple images are identified both
 photometrically and geometrically.  Improving on the CLASH program,
 the current FF program is destined to leave a legacy of exquisite
 quality data on gravitational lenses. The superior depth attained by
 the FF program, promises a wealth of faint multiply lensed galaxies,
 as the first examples are already showing
 (\cite{Zheng2014,Atek2014,Johnson2014,Richard2014,Jauzac2014}, Lam et
 al. 2014 in preparation).

 The cluster A2744 was the first one being released within the HFF
 program. More recently, new data from the FF was made available on
 the cluster MACSJ0416.1-2403, a massive merging galaxy
 cluster at redshift $z=0.4$.  A first analysis based on CLASH data
 on MACSJ0416 was performed by \cite{Zitrin2013} (Z13 hereafter).  In
 that paper the authors identified 70 multiple images from 23
 background galaxies. The strong lensing analysis of Z13 reveals a
 very elongated structure for the dark matter distribution. The
 addition of the new data from the FF program helps in clarifying some
 of the systems pblished in Z13.  In a very recent paper,
 \cite{Jauzac2014} (J14 hereafter), published a parametric solution
 and extends the original set of system candidates of Z13 to more than
 50 systems.  Also in two additional very recent papers, six of the FF
 clusters are studied using strong lensing
\cite{Johnson2014}, or a joint strong-weak lensing analysis \cite{Richard2014}. Among 
these clusters, the cluster MACSJ0416 is also included. A detailed
comparisson of our results with those derived using parametric methods
is beyond the scope of this paper and will be considered in a future
paper.

 Colliding clusters are of special interest for understanding the
 properties of DM partciles and the dynamics of the infalling systems
 through hydrodymanical modelling such as that of the bullet cluster
 \cite{Springel2007,Mastropietro2008} and more recently for the
 bulet like 'El Gordo' system, \cite{Molnar2014}. In
 particular, colliding clusters have been used to set stringent limits
 on the scattering cross section (see \cite{Kahlhoefer2014} for a
 recent discussion).  In \cite{Kahlhoefer2014} the authors conclude
 that small separations between the DM and the luminous matter
 (galaxies) are only possible right after the collision but not on
 more relaxed systems. If MACSJ0416 is in the turnaround phase (as suggested in cite{Mann2012}) 
 we should not expect a separation between the DM and the luminous matter (galaxies)
 for values of the dark matter cross section $\sigma/m \sim 1 cm^2/g$
 or less.  In \cite{deLaix1995,Spergel2000} they discuss how
 self-interacting dark matter can explain the flattening of cluster
 cores. In addition, \cite{Rocha2013} study simulations of
 self-interacting DM and derive some interesting conclusions that will
 be discussed later.

 It is possible to derive constraints on the DM self-interaction
 cross-section from the fact that such systems have survived the
 collision.  If self-interaction occurs between DM particles, one
 would expect a shift between the positions of the DM peak and galaxy
 positions \cite{Markevitch2004}. A detailed study of the mass
 distribution responsible of the lensing distortions around the
 central galaxies in MACSJ0416 has the potential to reveal such
 shifts. The pronounced elongation of MACSJ0416 might help in
 distinguising this possible shift.

\begin{figure*}  
\centerline{ \includegraphics[width= 15cm]{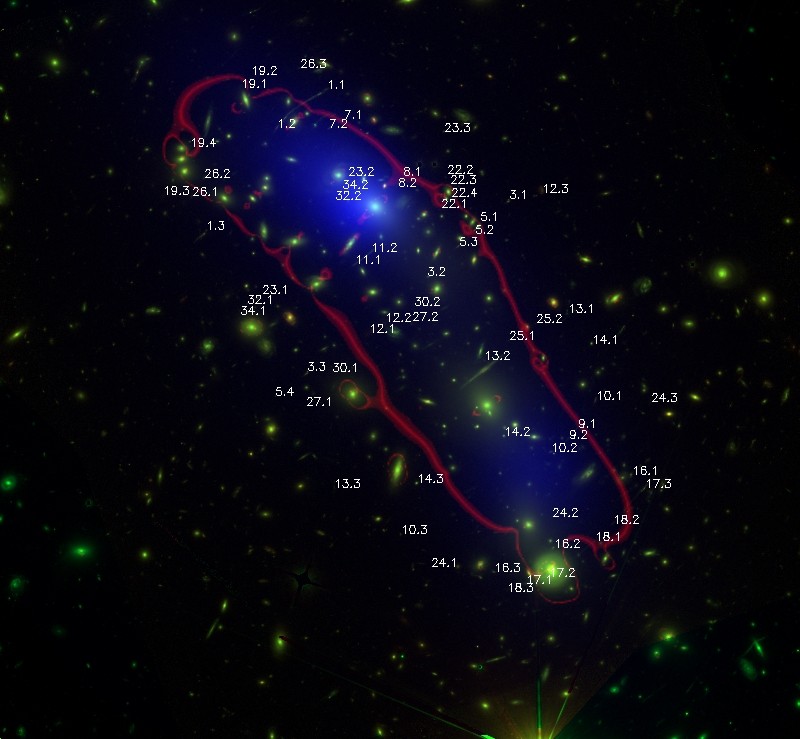}}          
   \caption{Relative position of the systems used in the reconstruction (part I). 
            The critical curve corresponds to the case i) described in \ref{sect_fid_gal}. 
            The two blue clouds correspond to the peaks of the X-ray emission from Chandra. 
            The field of view is 2.4 arcminutes.} 
   \label{fig_SystemsI}  
\end{figure*}

 In a recent paper based on X-ray and optical data, \cite{Mann2012}
 clasifies the cluster MACSJ0416 as a clear example of a
 post-collision merger and they add that the cluster could be even in
 the phase after turnaround. However, due to relatively poor X-ray
 data, the pre-collision scenario can not be ruled.  Deeper
 observations in X-rays could confirm the premerger scenario if an
 enhancement in the temperature is observed between the two
 subclusters, \cite{Ricker2001}.  Chandra data reveals an offset
 between the peaks in the X-ray emissivity and the cD galaxies,
 \cite{Mann2012}, which is consistent with the post-merger scenario
 (see also figure \ref{fig_SystemsI}).  The fact that one of the X-ray
 peaks is ahead of one the main cD galaxies would be consistent with
 the turnaround scenario.

 Through the paper we assume a cosmological model with $\Omega_M=0.3$, $\Lambda=0.7$, 
 $h=h_{100}= 100 km/s/Mpc$. For this model, 1 arcscec equals $3.87 kpc/h_{100}$ at the distance of the cluster. 

\begin{figure*}  
\centerline{ \includegraphics[width= 15cm]{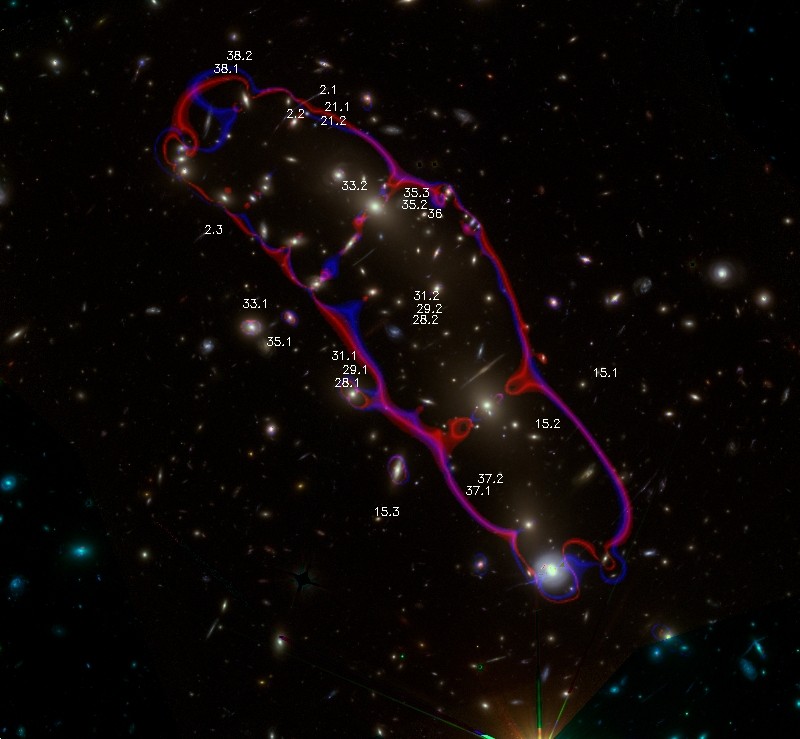}} 
   \caption{Relative position of the systems used in the reconstruction (part II). In this case,  
            the critical curves ($z_s=3$) are for for the cases ii) (red) and 
            iii) (blue) in section \ref{sect_fid_gal}.}
   \label{fig_SystemsII}  
\end{figure*}  

\section{FF data}\label{sect_FF_data}

In this paper we used public imaging data obtained from the ACS (filters: F450W, F606W and F814W) 
and the WFC3 (F105W, F125W, F140W and F160W), retrieved from the Mikulski Archive for Space 
Telescope (MAST). Some of these bands where not available at the time when the analaysis of Z13 was made. 
The new color bands are useful to identify systems that match (or do not match) in color but also to go 
deeper in magnitude. New observations on this cluster are planned for the near future in the optical bands. 
The data used in this paper consists of 50\% of the data that will be available in the near future. 

We combined all the filters into three color bands to produce the color image of the new 
candidate systems shown in the Appendix. The image was processed in Fourier space to reduce 
the contribution of large angular scale modes due to diffuse light from the member galaxies and later combined 
to produce a color image. The Fourier filtering introduces artificial changes in the brightness of sources around 
bright sources (galaxies and stars). The different angular resolutions of the optical and IR bands introduces also 
an element of uncertainty when processing the images in Fourier space so images that fall near bright sources may 
have their colors slightly distorted. We have checked that the colors of our candidates agree in the original 
color image (that does not include the filtering in Fourier space).

\section{Lensing data}\label{sect_lensing_data}

Our strong lensing data set is primary based on the system identification of Z13. Some of the photometric 
redshifts in that work have been updated with the recently published ones in J14 (that appeared at the time when 
we where finzalyzing this paper). Although the compilation of J14 superseeds the systems published in Z13, 
we rely on the early identification of Z13 and leave for a different paper the analysis based on the systems of 
J14 (some of which are more reliable than others). 
In that work, the authors extend the original set of 23 systems in Z13 to over 50 candidate systems. 
A proper comparisson of our results and theirs will be published in a separate paper and once the full FF 
data set is available. However, in section \ref{sect_35} we make a small comparisson in one of the new systems 
that is common to both data sets. Despite the fact that the system used in our work and theirs is different, 
a simple consistency test shows a high degree of agreement between our solution and theirs.

Our strong lensing data set is listed in table A1 in the appendix. Systems 1 through 
23 are taken directly from Z13 and systems 24 through 38 are new candidates 
identified in this work. 
Systems 6 and 20 in Z13 are dubious and have been removed after color 
comparisson with the new addition of the IR data from the FF program. Some of the counterimages 
in Z13 have deliverately not being used in this work as the proximity of multiple candidates makes it 
difficult to secure the identification. However we should note that most of these counterimages are in 
general in good agreement with our lens model so their inclusion in our analysis should have a small impact in 
the reconstructed mass. A few of the arclets in Z13 have been rematched, in particular  
12.3 (we use a different candidate from Z13 at one of their predicted positions) 
15.3, 19.4 and 22.4. The new selection of systems is based on a combination of color matching 
(after including the new data from the FF program) and comparisson with a lens model. The lens model used 
for this comparisson is obtained after using the 23 original systems of Z13 and our free-form algorithm. 
The lens model allows us also to identify new system candidates in the new optical+IR data by predicting the 
positions where counterimages are expected (for a given redshift) and also by predicting the right orientation and 
morphology of the lensed images. A few examples showing relensed images and involving systems 1,2,3 and 4 
are shown in the appendix in figure \ref{fig_Relensed_1to4}. The relensed image of the special new system 35 is shown 
(and discussed) also in section \ref{sect_35}. 
    
Stamps from the complete data set in table A1 are also shown in the appendix\footnote{also available 
in this website http://www.ifca.unican.es/users/jdiego/MACS0416\_March2014/Systems/}.
The relative positions of the images defining our lensing data set is shown in figures \ref{fig_SystemsI}  
and \ref{fig_SystemsII}. The data set has been divided into two groups for clarity purposes. 
Figure \ref{fig_SystemsI} shows part of the data set together with one of the critical curves derived in this work 
(see section \ref{sect_fid_gal}). 
In blue color, we also show an adaptively smoothed version of the 
X-ray emission from Chandra data\footnote{ivo://ADS/Sa.CXO\#obs/10446} for reference. 
The Chandra data has been smoothed using the code {\small ASMOOTH}, \cite{Ebeling2006}. 
Figure \ref{fig_SystemsII} shows the remaining systems together with two critical curves corresponding to 
two alternative solutions (see section \ref{sect_fid_gal})

\section{Reconstruction method}\label{sect_method}

We use the improved method, WSLAP+, to perform the mass reconstruction. 
The reader can find the details of the method in our previous papers
\citep{Diego05a,Diego05b,Diego2007,Ponente2011,Sendra2014,Diego2014}. Here we
give a brief summary of the most essential elements. \\

Given the standard lens equation, 
\begin{equation} \beta = \theta -
\alpha(\theta,\Sigma(\theta)), 
\label{eq_lens} 
\end{equation} 
where $\theta$ is the observed position of the source, $\alpha$ is the
deflection angle, $\Sigma(\theta)$ is the surface mass density of the
cluster at the position $\theta$, and $\beta$ is the position of
the background source.  Both the strong lensing and weak lensing
observables can be expressed in terms of derivatives of the lensing
potential. 
\begin{equation}
\label{2-dim_potential} 
\psi(\theta) = \frac{4 G D_{l}D_{ls}}{c^2 D_{s}} \int d^2\theta'
\Sigma(\theta')ln(|\theta - \theta'|), \label{eq_psi} 
\end{equation}

where $D_l$, $D_{ls}$ and $D_s$ are the
angular diameter distances to the lens, from the lens to the source
and to the source, respectively. The unknowns of the lensing
problem are in general the surface mass density and the positions of
the background sources. As shown in \cite{Diego2007}, the
strong lensing problem can be expressed as a system of linear
equations that can be represented in a compact form, 
\begin{equation}
\Theta = \Gamma X, 
\label{eq_lens_system} 
\end{equation} 
where the measured strong lensing observables are contained in the
array $\Theta$ of dimension $N_{\Theta }=2N_{SL}$, the
unknown surface mass density and source positions are in the array $X$
of dimension $N_X=N_c + N_g + 2N_s$ and the matrix $\Gamma$ is known
(for a given grid configuration and fiducial galaxy deflection field) 
and has dimension $N_{\Theta }\times N_X$.  $N_{SL}$ is the number
of strong lensing observables (each one contributing with two constraints,
$x$, and $y$), $N_c$ is the number of grid points (or cells) that we use to divide
the field of view. In this paper we consider a regular grid of $N_c=32\times32=1024$ cells 
covering the field of view shown in figures \ref{fig_SystemsII} and \ref{fig_SystemsII} (2.4 arcminutes).
$N_g$ is the number of deflection fields (from cluster members) that we consider.  
In this work we test three different configurations for 
the defelection field where  $N_g$ is equal to 1 (all member galaxies conform a unique 
deflection field) or $N_g=3$ which corresponds to the case where the two main cD galaxies are 
considered independently from the rest of the cluster members. See section \ref{sect_fid_gal} for details 
of the three configurations. 
$N_s$ is the number of background sources (each contributes with two unknowns, 
$\beta_x$, and $\beta_y$, see \cite{Sendra2014} for details). The solution is found after
minimizing a quadratic function that estimates the solution of the
system of equations \ref{eq_lens_system}.  For this minimization we
use a quadratic algorithm which is optimized for solutions with the
constraint that the solution, $X$, must be positive. 
Imposing the constrain $X>0$ also helps in regularizing 
the solution as it avoids large negative and positive contiguous fluctuations.

\begin{figure*}  
   \includegraphics[width=11cm]{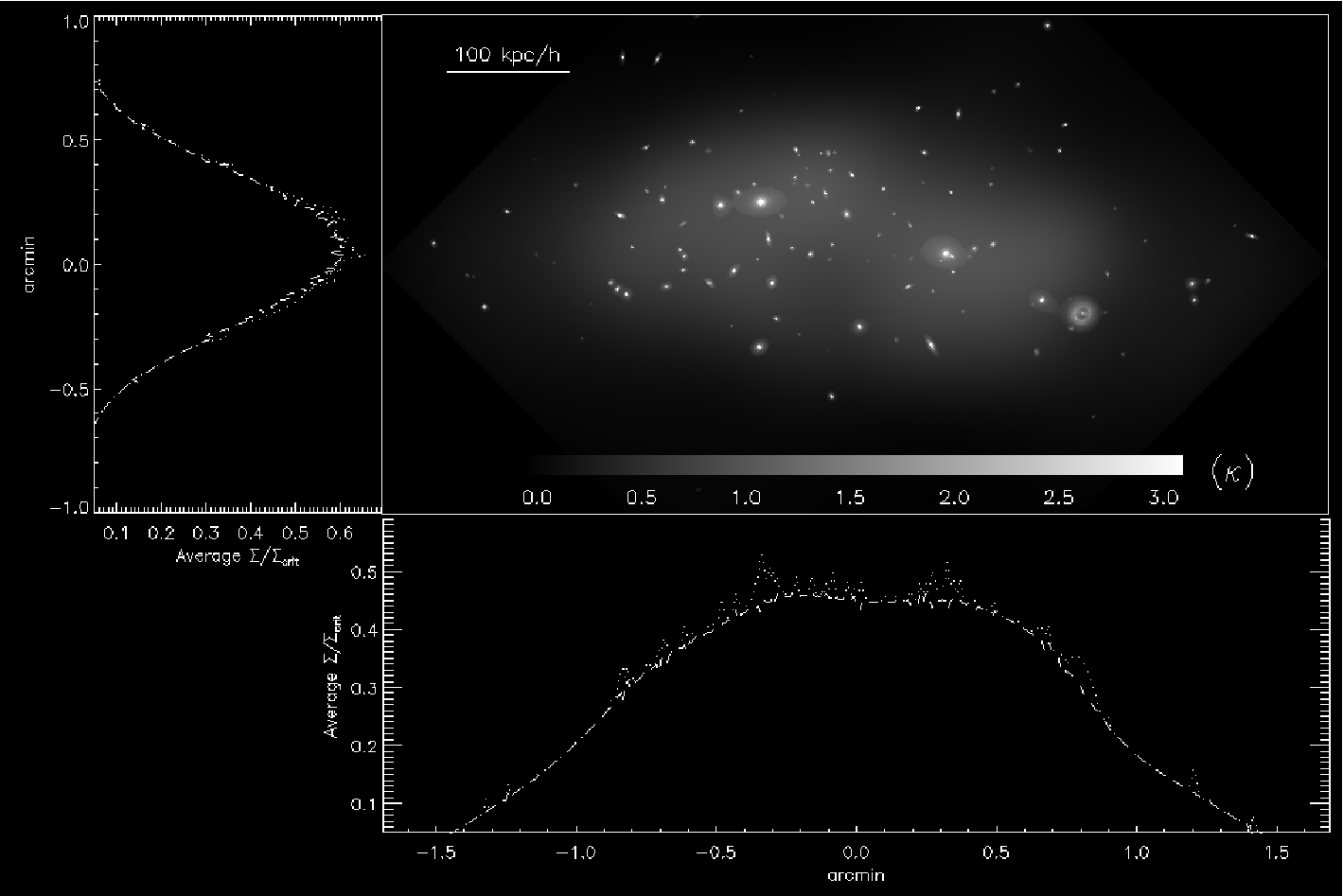} 
   \includegraphics[width=11cm]{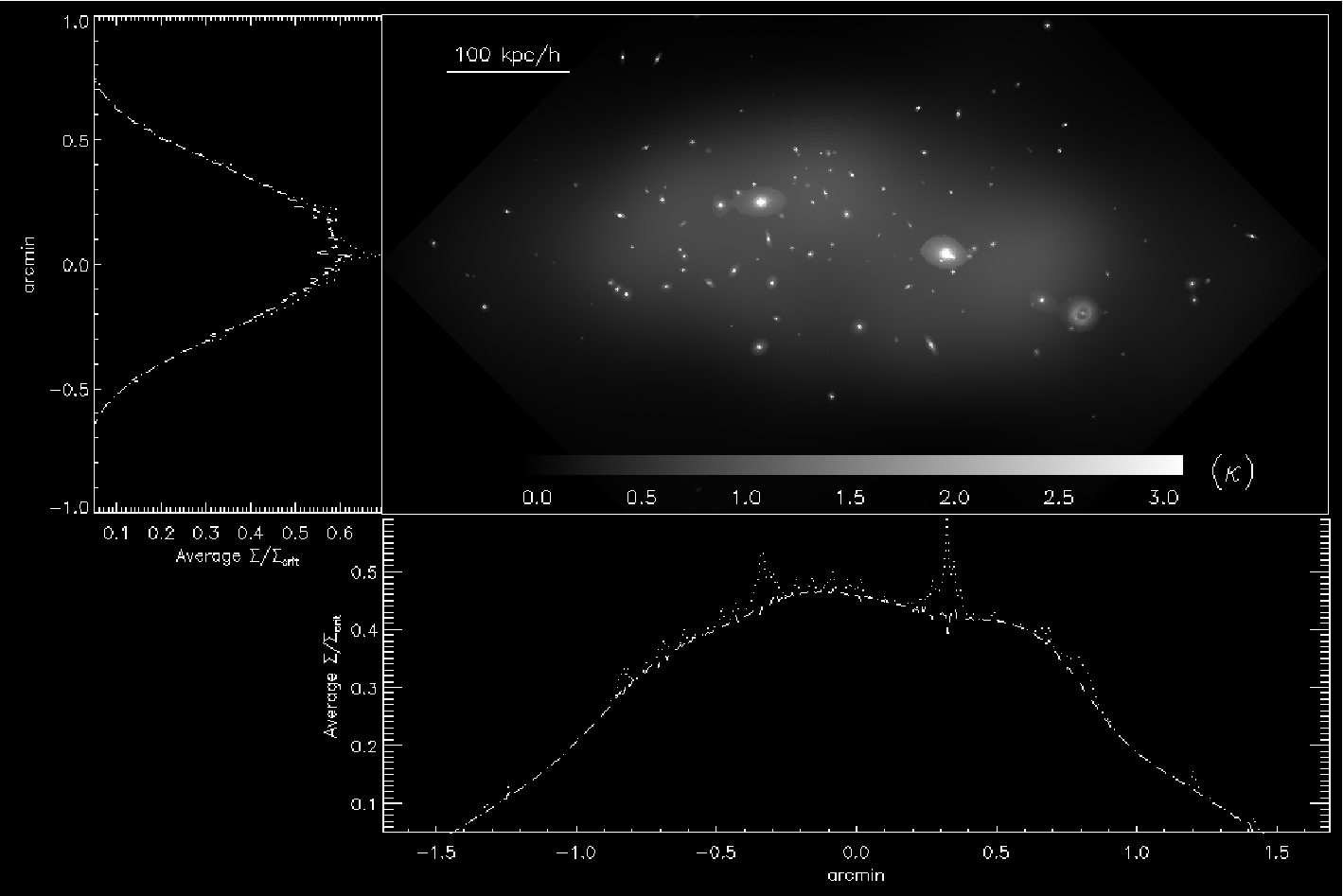} 
   \includegraphics[width=11cm]{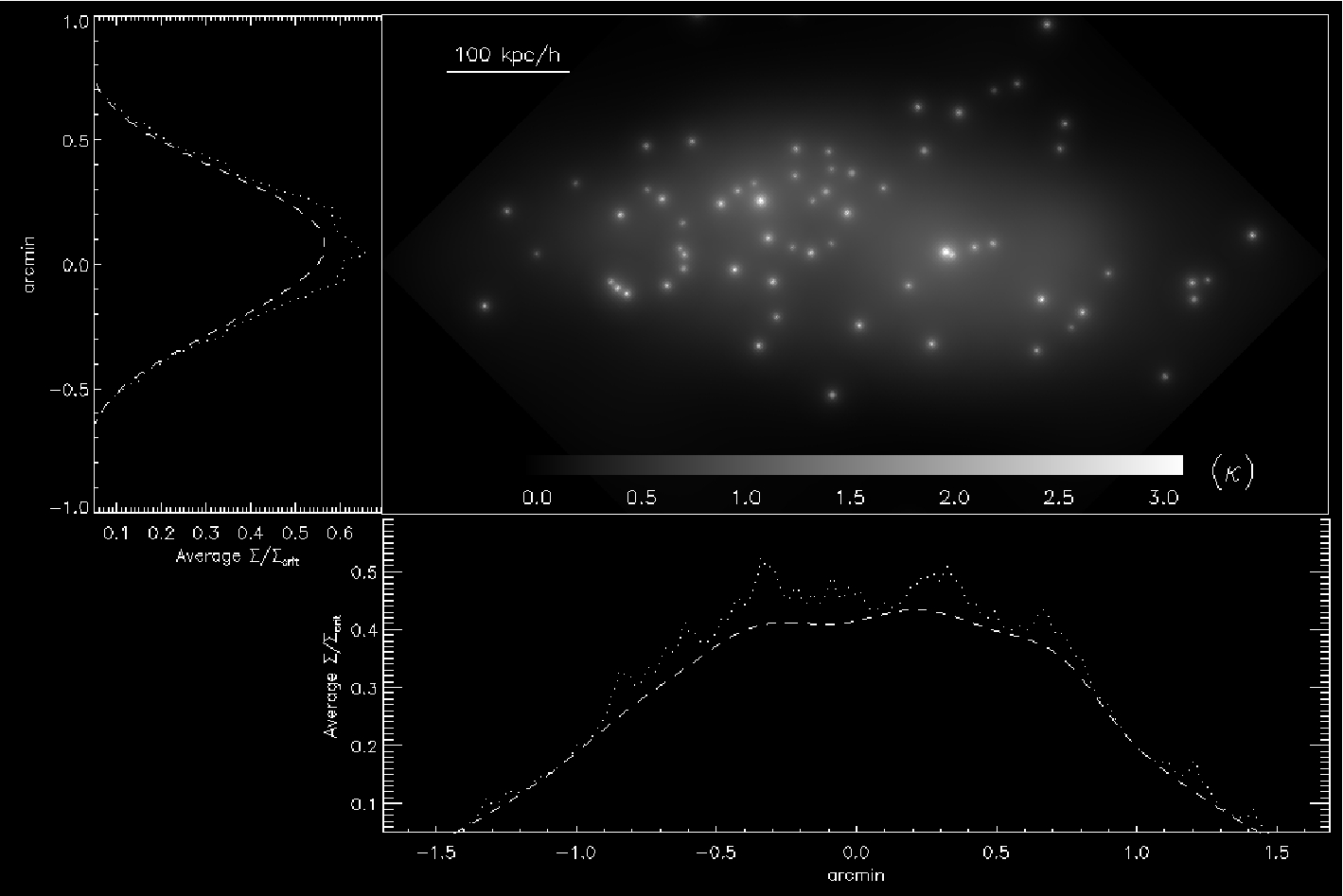} 
   \caption{Reconstructed mass in MACSJ0416 in units of its convergence (for $z_s=3$).
            2-dimensional convergence ($\kappa$) map and the mean value of the projected convergence along the 
            y and x axis respectively. The two curves correspond to the total mass (dotted) and to the grid-only 
            mass (dashed). 
            The convergence maps have been saturated beyond k=3 for clarity purposes. 
            The top plot is for case i) described in section \ref{sect_fid_gal}.
	    The middle plot is for case ii). 
            The bottom plot is for case iii).}
    \label{fig_Mass_profiles}  
\end{figure*}

\begin{figure*}  
   \includegraphics[width=5.5cm]{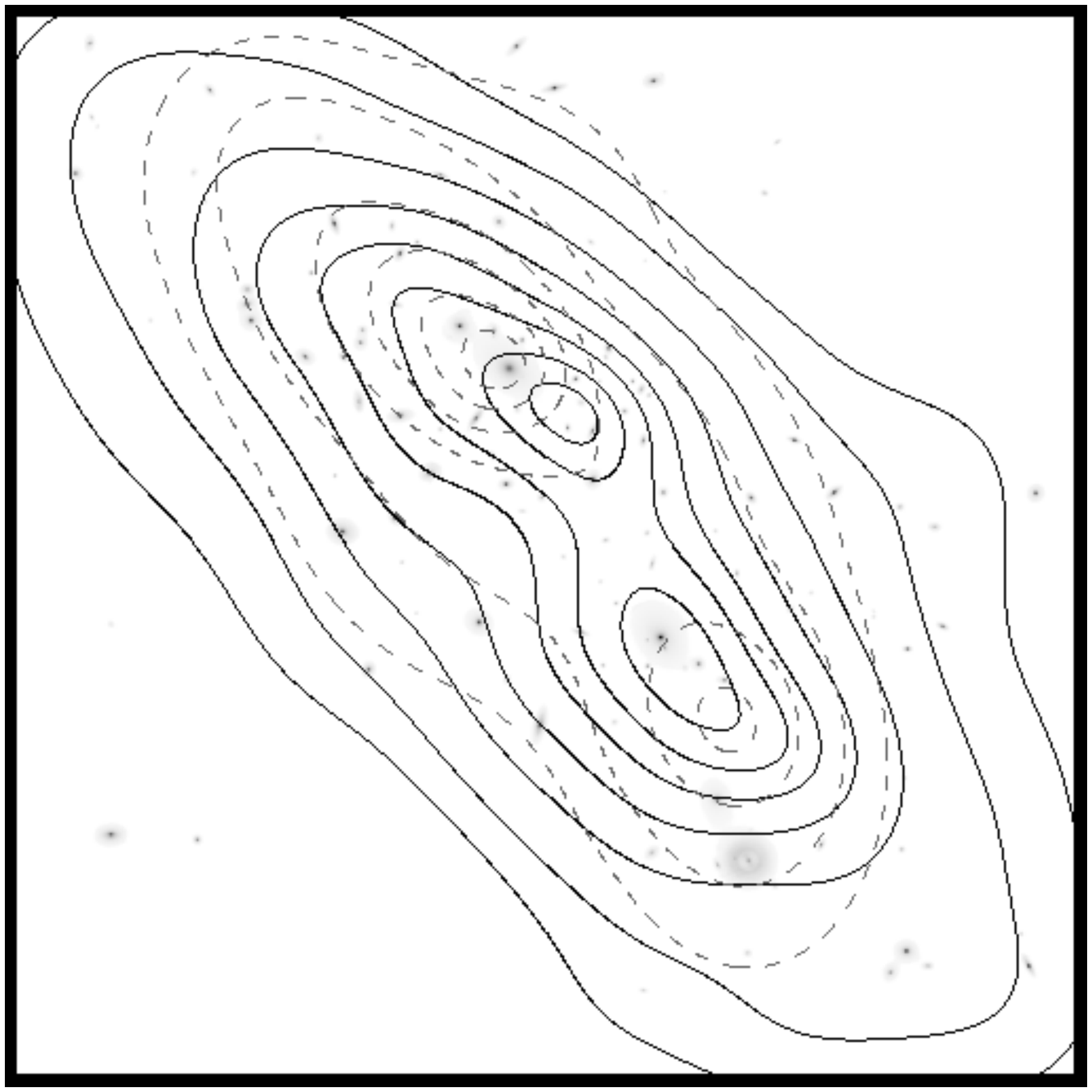} 
   \includegraphics[width=5.5cm]{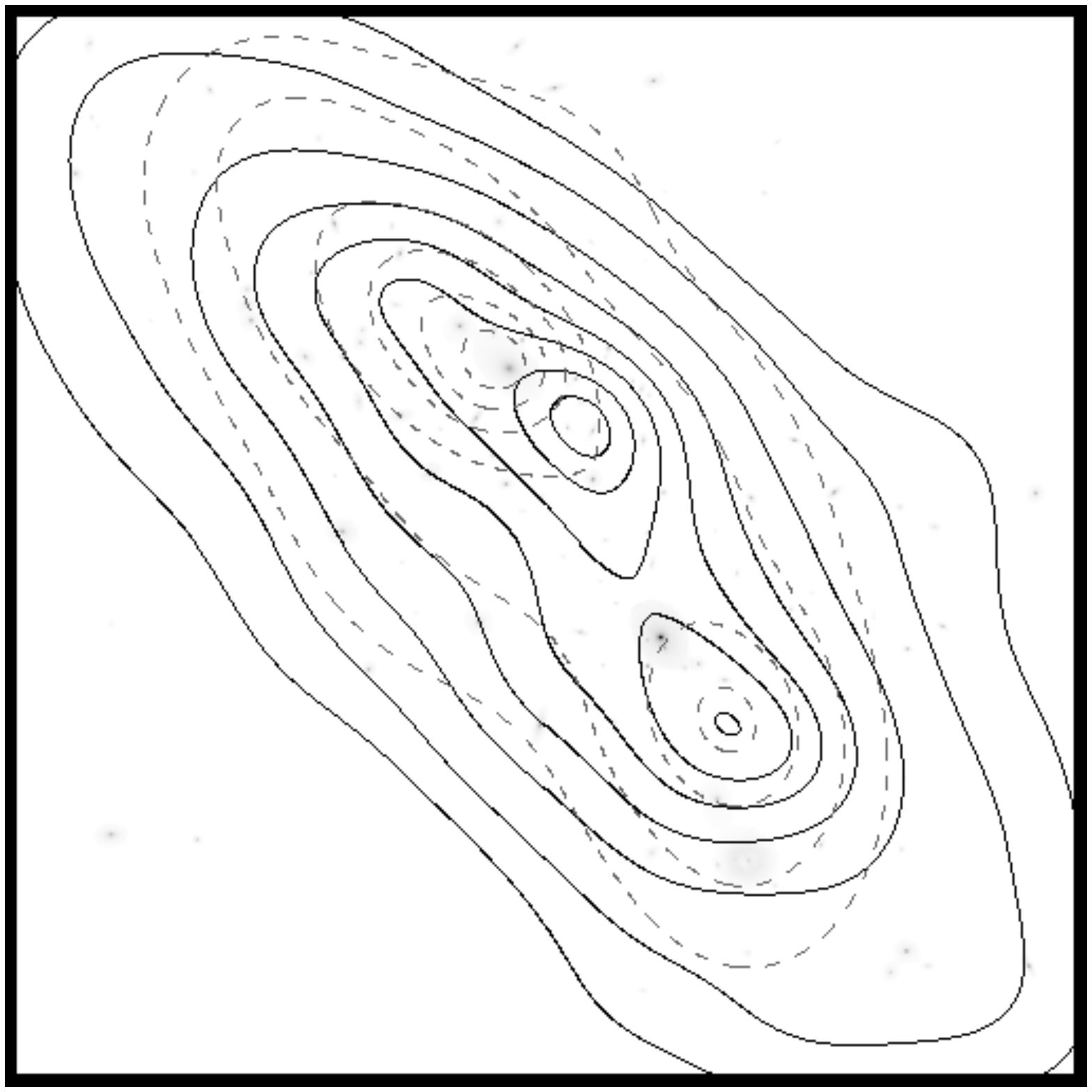} 
   \includegraphics[width=5.5cm]{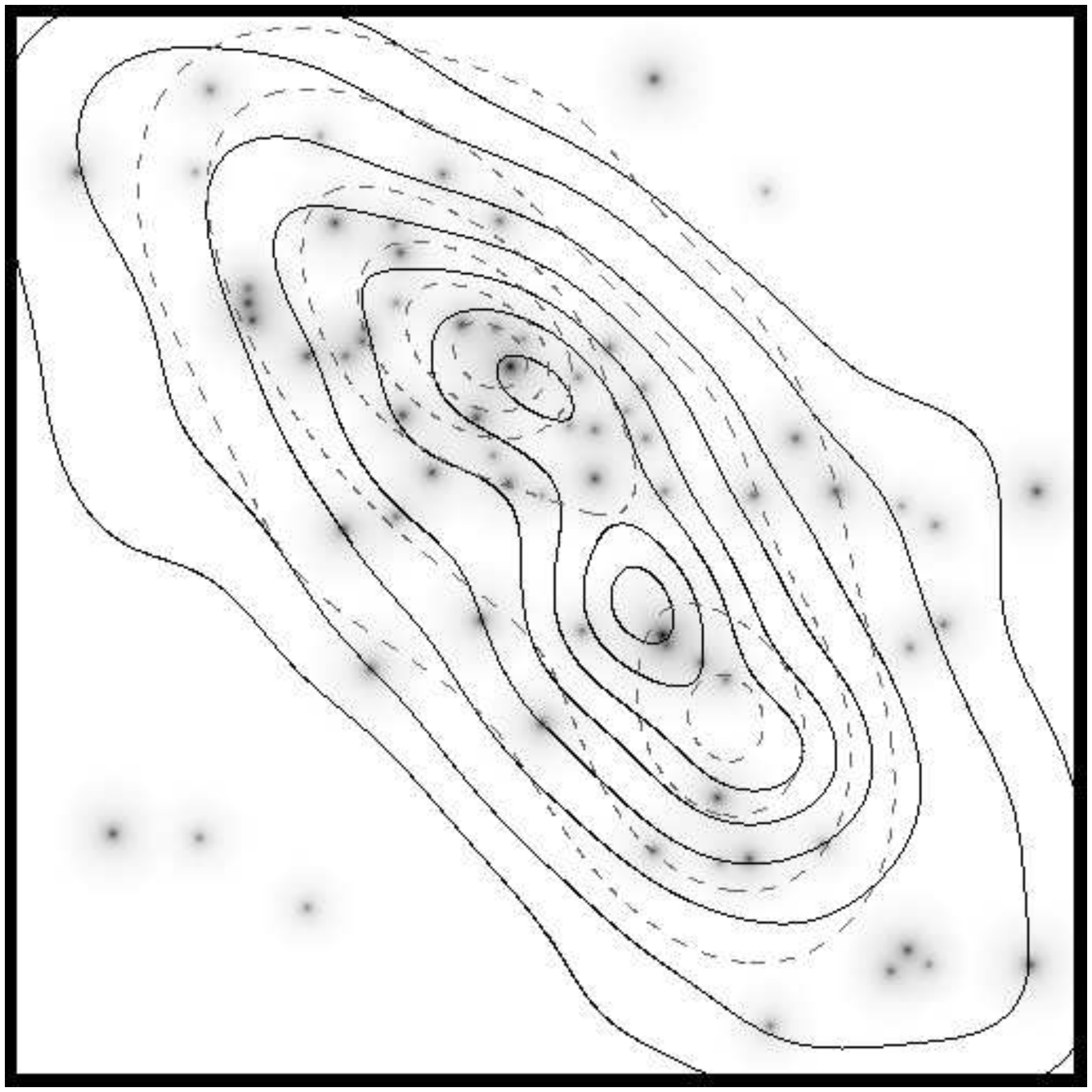} 
   \caption{Contours of the reconstructed mass of the grid component compared with the fiducial galaxy field. 
            From left to right, cases i), ii) and iii) respectively and described in section \ref{sect_fid_gal}. 
            In all cases, the contours correspond to 0.1,0.2,0.4,0.6,0.75,0.85,0.93,and 0.98 times the 
            maximum of the soft component. In all panels, the dashed contours shows the smoothed X-ray emission 
            from Chandra.
            }
    \label{fig_Mass_Contour}  
\end{figure*}

\subsection{Fiducial galaxies}\label{sect_fid_gal}
The member galaxies defining the fiducial field are all elliptical galaxies selected from the red sequence 
plus two very luminous (and possibly massive) spiral galaxies that lay near the critical curves. 
The selected galaxies are shown in figure \ref{fig_Mass_Contour}.
In the reconstruction, we need to assume some masses for the fiducial field (cluster members). 
We explore three different alternatives. \\

\noindent
i) The light of the galaxies traces their mass (LTM). We select the member galaxies from the red sequence and 
   make their masses proportional to the optical flux in the F814w band. The total mass of the galaxy field is 
   then normalized to have $7.0 \times 10^{13} M_{\odot}/h$. In this case there is only on parameter, $C_1$, 
   in the vector $X$ that re-scales the entire fiducial field.\\
ii) Like the above but the two main cD galaxies have now their own parameter C, hence we have three parameters 
    $C_1$ for the non-cD galaxies, $C_2$ for the NE cD galaxy and $C_3$ for the SW galaxy. \\
iii) Instead of assuming that the mass follows the light, we associate a circularly simmetric NFW profile 
     to each galaxy, \cite{NFW1997}. 
     The masses take different values than the ones in cases i) and ii) and also we use 3 
     fiducial fields like in case ii). 

The fiducial masses for these three cases are shown more explicitely in figure \ref{fig_Mass_Contour} below.

\section{Dark matter distribution}\label{sect_results}

The mass reconstruction is shown in figure \ref{fig_Mass_profiles} for the three cases discussed in 
section \ref{sect_fid_gal}. The field of view has been rotated by 45 degrees (counterclockwise) 
so the NE cluster is now at the left and the SW cluster is now to the right. 
Projections along the main axes in the cluster are shown in the left and bottom panels for each case. 

A striking result is the relatively high symmetry between both subclusters suggesting a mass ratio 
of 1:1 for this merger. From the morphological point of view, 
both subclusters seem to have similar masses and morphologies when projected in the two orthogonal directions. 
The extended soft dark matter halo is divided in two sub-halos, each associated with one of the two 
cD galaxies in the cluster. These subhaloes do not show a peculiar 
symmetry, in particular there is no clear evidence for the elliptical spheroids assumed in parametric methods but 
rather the soft dark matter halo shows a rather irregular shape.  
When comparing the 3 reconstructions, we find that the assumptions made on the fiducial field play a secondary 
but not entirely negligible role. Comparing cases i) and ii), the solution seems to prefer the two main cD galaxies 
having larger mass associated to their halos. The same result is observed when all the haloes are assumed to 
follow a (symmetric) NFW profile as seen in case iii). 
It can be also observed, that when the galaxies take more 
mass this is at the expense of the mass in the grid component, maintaining the total mass more or less constant. 
This type of degeneracy is expected in this particular cluster due to the relatively small number of radial arcs 
that fall close to the main cD galaxies. The mass distribution around these galaxies remains then relatively 
unconstrained. 

Figure \ref{fig_Mass_Contour} shows the contours of the soft component (grid only) 
superimpossed on the fiducial galaxies for the 3 cases described in subsection \ref{sect_fid_gal}. 
The three cases reproduce almost identical results around the position of the critical curves. 
However, small differences can be appreciated in the central region with the 
models showing some dependency with the assumptions made about the central cD. Interestingly, a shift can be 
appreciated between the position of the cD galaxies and the peak of the diffuse component when we assume that 
the mass in the galaxies traces their light, case i). 
The shift is more pronounced when the two cD galaxies are unlocked from the remaining 
galaxies, case ii). When the two cD galaxies are locked to the other galaxies, case i), the grid takes 
on the possible missing mass around these two galaxies shifting the peak of the diffuse mass closer to the 
position of the galxies. This is made more evident in figure \ref{fig_Mass_profiles} above, where in case ii) we 
appreciate a higher contribution from the main cD to the cluster mass than in case i). In case iii) we do not 
observe a significant shift between the peaks of the soft dark matter halos and the 2 main galaxies. However, 
in this case, the more extended nature of the fiducial galaxies reduce the relative importance of the grid 
as the superposition of the extended haloes in the fiducial galaxies can reproduce some of the soft features 
that otherwise would be reproduced by the grid. Case iii) can be 
considered as a case where the galaxies compete more with the grid than in cases i) or ii) or more mathematically, 
the base defined by the grid+galaxies is less orthogonal than the other two cases. 
When comparing the soft DM component with the X-ray emission, there are small offsets between the two components 
but these offsetys depend on the assumptions made for the fiducial galaxies. Also, the peak in the X-ray emission 
is now ell defined and depends on the resolution at which the X-rays are smoothed. Deeper X-ray observations are 
needed to better constrain the position of the X-ray peaks.

The fact that the DM peak from the soft dark matter halo (grid) may be shifted with  
respect to the position of the cD galaxies is an interesting feature. 
Since the galaxies can be trully considered as collisionless, the DM 
could have its peak shifted with respect to the main galaxies if they have a significant cross section, $\sigma/m$. 
This case has been recently studied by \cite{Kahlhoefer2014} where they consider a value of 
$\sigma/m \approx 1 cm^2/g$. They find that small shifts are expected only inmediately after the collision between 
the two clusters. The peaks of the DM and the dominant galaxy quickly converge to the same point shortly after the 
collision. However, they also find that the baryonic and the DM particles exhibit different profiles with the 
galaxies showing a tail moving ahead (in the direction of the movement of the halo) of the DM halo and the DM 
showing a more elongated tail in the opposite direction of the movement. 

The contours in figure \ref{fig_Mass_Contour} show a clear elongation of the DM haloes in the axis of the 
merger. Interestingly, the halo in the SW shows a more pronounced elongation. This picture is 
consistent with the situation discussed in \cite{Kahlhoefer2014} if a non-negligible value for  
$\sigma/m \approx 1.5 h^{-1} cm^2/g$ is assumed.  

Another possibility is that the elongation may be due to our reconstructed mass distribution  
being also sensitive to the projected mass of the plasma. If the plasma is displaced with respect to the DM, this 
could produce an elongation in the projected mass distribution in the direction of the displacement. This possibility 
will be discussed in more detail in section \ref{sect_dynamics}. The fact that the elongation traces the shape of 
the maxima of the X-ray contours suggests that this scenario is possible. In Lan et al. (2014), a similar excess of mass 
not directly linked to the observed galaxies is also found in a region with significant X-ray emission.  

Regarding the masses contained in the galactic component, for case i), the derived total 
mass of the fiducial galaxies is $8.21 \times 10^{12} M_{\odot}/h$.   
For case ii), the derived masses of the two main cD galaxies are 
$1.13 \times 10^{12} M_{\odot}/h$ for the NE cD galaxy and $1.95 \times 10^{12} M_{\odot}/h$ 
for the SW cD galaxy. The combined total mass of the remaining fiducial galaxies equals 
$5.36 \times 10^{12} M_{\odot}/h$.  
For case iii), the derived masses of the two main cD galaxies are 
$1.41 \times 10^{12} M_{\odot}/h$ for the NE cD galaxy and $1.34 \times 10^{12} M_{\odot}/h$ 
for the SW cD galaxy. 
The combined total mass of the remaining fiducial galaxies equals 
$2.08 \times 10^{13} M_{\odot}/h$. The higher galaxy masses in case iii) can be attributed to their larger 
angular size that takes away some of the mass from the grid compondent.

\begin{figure}  
   \includegraphics[width=8cm]{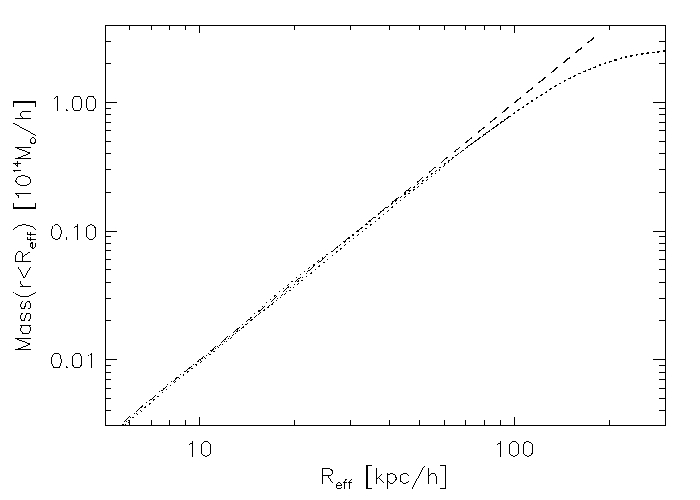} 
   \caption{The dotted line corresponds to the integrated mass as a function of effective radius 
            (see text). The dashed line is a power law, $M \propto R_{eff}^2$
            }
    \label{fig_IntegratedMass}  
\end{figure}  

In order to get an estimate of the integrated mass as a function of 
radius, and given the high assimetry of this cluster, we integrate the masses following the contours of the 
cluster. Using the contours (similar to the ones shown in figure \ref{fig_Mass_Contour}) we can define an 
effective radius, $R_{eff}= \sqrt{A/\pi}$, where $A$ is the area enclosed by a given contour. By setting 
different thresholds in the projected mass maps we can build a set of contours, each with a given enclosed 
area, $A$, and  $R_{eff}$. The integrated mass is shown in figure \ref{fig_IntegratedMass} as a function of $R_{eff}$. 
The mass in the y-axis corresponds to the combined total mass of the two 
sub-clusters above a given threshold (with an associated radius, $R_{eff}$). In figure \ref{fig_IntegratedMass} 
we also show a power law, $M(<R_{eff}) = 10^{-4} (R_{eff}/h^{-1}kpc)^2 M_{\odot}/h$ that fits nicely the integrated 
mass in the regime $5-50$ kpc/h. A mass growing as the square of the radius is expected for a surface mass 
density that is constant with radius. The nearly-flat behaviour of the surface mass density in the region  
$5-50$ kpc/h is made more evident when we compute the radial profiles of each subcluster after masking the 
other one out. 

\begin{figure}  
   \includegraphics[width=8cm]{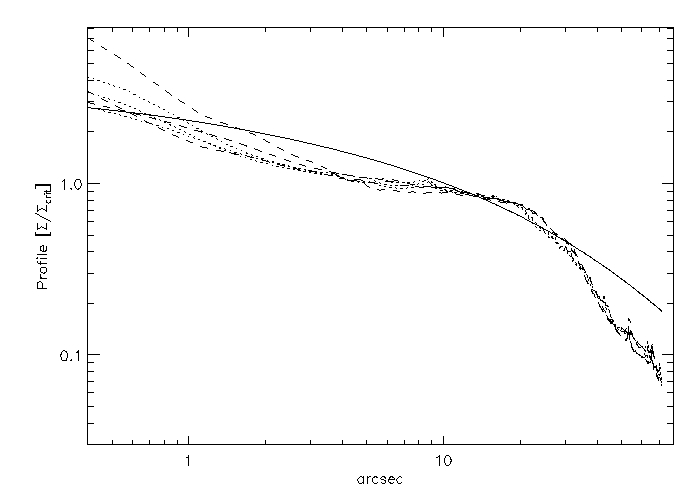} 
   \caption{Profiles of the two cluster haloes for the three different cases. 
            The profiles are centered in the two main cD galaxies. 
            The NE halo corresponds to the dotted lines and the 
            SW halo corresponds to the dashed lines. 
            In the computation of the profiles, we have masked the other half of the cluster.
            The solid line corresponds to a projected NFW model with a truncation radius of $R_{200} = 2.5$ 
            Mpc/h and $C=10$.}
    \label{fig_Profiles}  
\end{figure}

Figure \ref{fig_Profiles} shows the circularly-averaged profiles for each subcluster in the 3 cases discussed above. 
The centre of the cluster is the corresponding cD galaxy. The dotted line is for the NE cluster and the 
dashed line is for the SW cluster. The profiles for each subcluster are computed after excluding 
the other one from the calculation. The division between the two clusters (and the mask) is defined by a straight line 
going between the bottom left corner to the top-right corner of figure \ref{fig_Mass_Contour}.  
The profile is computed in units of the critical surface mass density assuming a redshift of $z_s=3$. The two 
profiles are almost identical, both in their amplitudes as well as in their slopes with the exception of the very 
central region where the cD profile dominates. 
The profiles beyond 30 arcsec ($115 kpc/h$) show a sharp decline. 
Part of this decline is due to the fact that the profile is an average of an elongated surface mass density 
distribution. 
Beyond 30 arcseconds, the profiles behave very differently in the two orthogonal directions as shown in 
figure \ref{fig_Mass_profiles}. 
Hence, we should expect the profile to retain some of the features in the the x and y projected profiles in 
figure \ref{fig_Mass_profiles}. However, is important to emphasize that the profiles in figure \ref{fig_Profiles} 
are derived for each subcluster while masking the other one so the highly assymetric distribution of the double 
cluster is somewhat attenuated. 
Another reason for the decline of the profile at large radii is the lack of constraints. 
As shown in our earlier works, our algorithm understimates the real dark matter distribution when going into 
the outer radii where there is no more strong lensing constraints. 
On the other extreme, the inner part of the profiles depend on the assumptions made in the profiles 
of the fiducial field for the member galaxies and it is affected by uncertainties on these profiles. 
The lack of constraints in the very central region limits also the accuracy at which this part of the profile 
can be estimated. However, the mass distribution in the range 5-30 arcseconds ($20-115 kpc/h$) can be trusted. 
The mass profiles in this range show little dependency with the assumptions made on the fiducial field. 
In figure  \ref{fig_Profiles} we also plot as a reference the projected surface mass density 
(in terms of the same critical surface mass density) of a NFW profile. 
A simple NFW profile can not fit the mass distribution of each subhalo. Modifications of 
the NFW profile, like a cored NFW profile would produce a better fit.

\section{The curious case of system 35}\label{sect_35}
\begin{figure*}  
   \includegraphics[width=15cm]{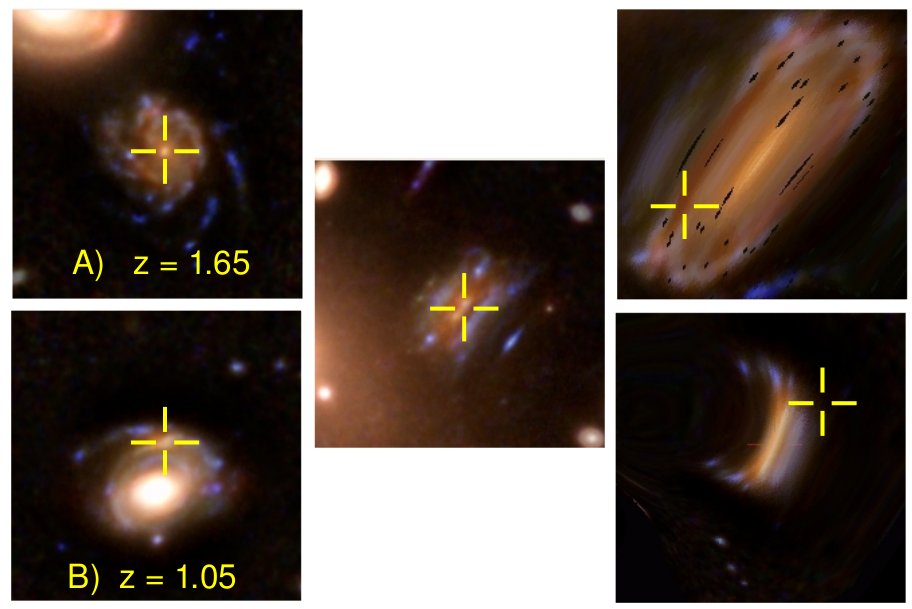} 
   \caption{The two possible configurations for system 35. 
            The two counterimages 35.2 and 35.3 are shown in the middle 
            column. The left column shows the two possibilities, A and B for the counter image 35.1, each one at a 
            different redshift. 
            The right column shows the predicted re-lensed image based on our model for case ii) and using the 
            corresponding delensed image on the left column. 
            The situation is similar for the models of case i) and iii). 
            The field of view is 7.2 arcseconds in all cases. The crosses mark the observed positions.}
    \label{fig_Syst35_AB}  
\end{figure*}
\begin{figure}  
  \begin{center}
   \includegraphics[width=5cm]{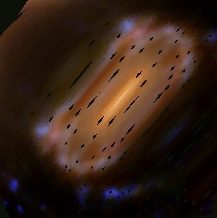} 
   \caption{As in figure \ref{fig_Syst35_AB}, predicted system 35 for option A in the case of a multiresolution 
            grid. The scale and position is the same as in the case of option A in figure \ref{fig_Syst35_AB}.}
    \label{fig_Syst35_MR}  
  \end{center}
\end{figure}

  
In this section we explore in more detail the new system 35. 
Despite being a relatively bright system with distintive 
features, it was not matched in the previous work of Z13. 
The reason maybe the fact that there are two possible images that could be the predicted third counterimage of 
this system. Usually, these kind of ambiguity is found in systems that exhibit none or few morpholical 
features, are unresolved or too faint. 
However, system 35 shows a relatively complex morphological and color-rich structure that should, in principle, 
make it relatively easy to identify its counterimage(s). 
This new system is also interesting because the two central counterimages are very close to the northern cD galaxy 
and they are sensitive to the density profile in the innermost part of the cluster. 
Images 35.2 and 35.3 are challenging to reproduce by lens models due to the fact that they are 
merging into a single image (or splitting from one single image) and hence are very sensitive 
to the small fluctuations in the gravitational potential at the centre but at the same time this offers 
a good opportunity to study in detail the potential surrounding the northern cD galaxy. 
In figure \ref{fig_Syst35_AB} we show the two possibilities for system 35, 
option A at redshift $z_s \approx 1.65$ and option B 
at redshift $z_s \approx 1.05$ (these redshifts are estimated from the lens model).  
The same plot shows also the predicted lensed images based on the two possibilities 
for the third counterimage. 
Both options reproduce the position, orientation and some of the color and morphological features 
of the observed images. 
The apparent size of the relensed image is better reproduced in the option B. On the other hand, 
option A contains a very compact nucleus that is more consistent with the observed double nucleus in 35.2 and 
35.3 while the nucleus appears more difuminated in option B. Based on the presence of the nucleus we adopt 
option A as the counterimage but noting that option B is equally valid (in terms of being fit by the model). 
Preliminary results (work in preparation) shows that the size is better reproduced with option A when we consider 
a multiresolution grid that increases the resolution of the matter distribution around the central NW 
cD galaxy (see figure \ref{fig_Syst35_MR}).

Interestingly, in the recent paper by J14, the authors consider option B in their analysis 
although they assume a slightly different redshift for the system ($z_s=1$). 
Although a more detailed and quantitative comparisson with J14 will be done in a subsequent paper, the fact that 
their alternative definition of system 35 (redefined as system 28 in their work) is coincident with the option B 
for system 35 discussed above, suggests that, at least in the regions of the cluster proved by system 35, 
the free-form model derived in this paper and the parametric model derived in J14 are consistent to a 
high degree. The ambiguity in system 35 can be eliminated once a spectroscopic redshift of the 
system 35 is obtained. The redshift will unambiguously determine which one of the options, A or B, is the 
correct one. This system is a good text book example of the ambiguity inherent to strong lensing where highly 
distorted images are difficult to match even if these images are bright and well resolved.

\section{Geometry and dynamics of the merger}\label{sect_dynamics}

Multiwavelength observations offers the opportunity to interpret the
observations in the context of the dynamical state and geometry of the
cluster. The fact that an offset is observed between the X-ray peaks
and the inferred local minima of the gravitational potential (at the
position of the cD galaxies) suggests that the plasma is being
affected by dynamical friction and is leaving the potential
well. These scenarios have been observed in different clusters with
the most famous being the {\it bullet} cluster. In these cases, the
plasma always lays behind the moving gravitational potential. In the
bullet cluster, the cluster is observed after the collision has
happened and the two clusters are moving in a plane that is close to
the plane of the sky so projection effects are small. In our case,
there is no obvious way to tel what the plane of the collision is by
looking at the DM and lensing maps. If we assume that the collision is
taking place in the plane of the sky, the SW clump shows the peak of
the X-rays displaced with respect o the DM peak but in the opposite
side that one would expect if a collision has already happened. If the
collision has not happened yet, and the two clusters are on course for
their first core passage, the X-ray peak in the SW would be on the right
side but the magnitude of the shift is too large since the plasma is
expected to dissociate from the DM at the time of closest approach.

\begin{figure}  
  \begin{center}
   \includegraphics[width=8cm]{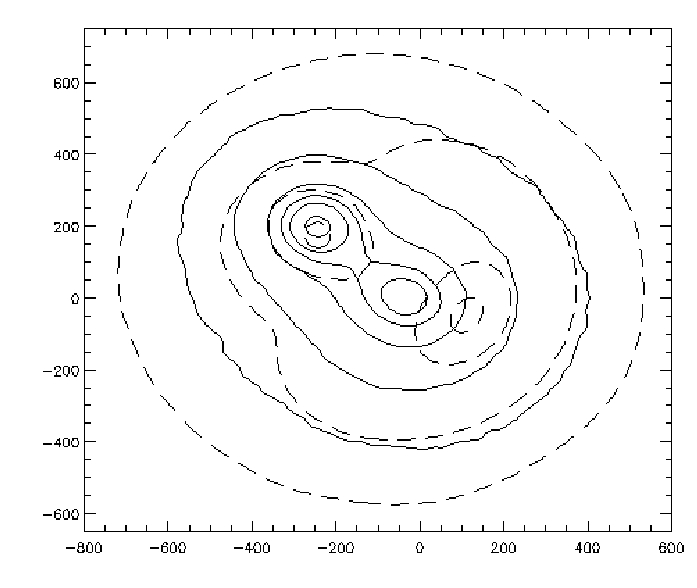} 
   \caption{Simulated cluster with the two clumps moving along the line of sight and the X-rays (dashed contours) 
            showing an offset with respect to the DM (solid contours) to the right of the south clump 
            due to projection effects. The axis are in kpc.}
    \label{fig_SimulCL}  
  \end{center}
\end{figure}

Alternatively, if the collision happens in a plane that is normal to
the plane of the sky, projection effects can help to explain the
relative positions of the DM and X-ray peaks.  
Figure \ref{fig_SimulCL} shows an example from a hydrodynamical simulation
that resembles the case of MACSJ0416.  Like in figure
\ref{fig_Mass_Contour}, solid contours correspond to the DM component
and dashed contours to the X-ray surface brightness. 
Our simulation is projected along an axis with 21 degrees inclination angle relative to the line of sight.  
The morphology is similar: the peaks of the dark matter and X-ray emission are almost in the same line. 
The dark matter peaks are located closer to each other. 
The large offset between the peak of the dark matter and X-ray emission of the infalling cluster 
is due to the ram pressure separating the gas form the dark matter component as it passes through the main cluster. 
With virial concentration parameters of 5 and 8, our simulation has smaller masses (2 and 0.5 $\times \rm M_\odot$) 
and a larger infall velocity (4500 km/sec) and produce a larger offset between the dark matter and the X-ray emission, 
therefore it is likely that a smaller infall velocity is necessary to produce the observed morphology.
The offsets between the peaks of the dark matter surface density and X-ray emission in the merging cluster are 
important in terms of understanding the physics of the collision (i.e., initial masses of the two components, 
the impact parameter and infall velocity). 

\begin{figure}  
  \begin{center}
   \includegraphics[width=9cm]{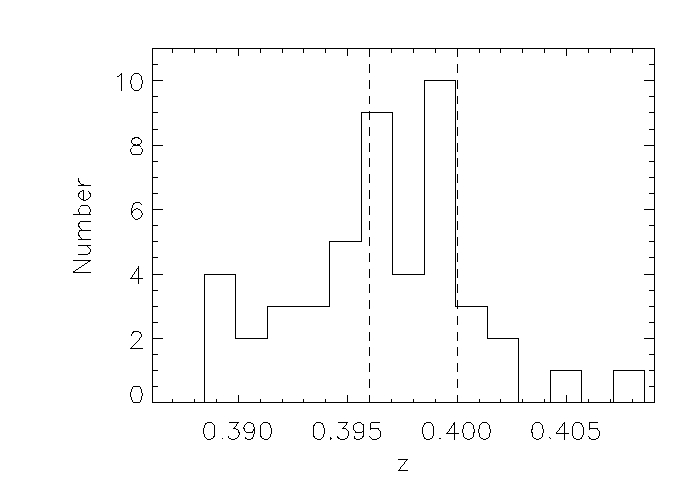} 
   \caption{Histogram of the galaxies with espectroscopic redshifts in the field (see Ebeling et al. 2014)}
    \label{fig_HistoZ}  
  \end{center}
\end{figure}  

In this simulation, the clusters are moving in a direction that is close to the line of sight
and due to the projection effects the X-ray peak in the SW seems to be
located ahead of the DM peak in the SW. However, the 3D simulation
shows that the X-ray peak is already behind the DM peak as
expected. The scenario where the collision happens in an axis close to
the line of sight is also supported by the difference in spectroscopic
redshift between the two cD galaxies. These redshifts where recently
published in \cite{Ebeling2014}. The NE cD galaxy has an spectroscopic
redshift of z=0.400 while the SW cD has an spectroscopic redshift of
z=0.396. 

The relative difference in redshift can be translated in a relative
velocity, $\delta v \approx 1200 km/s$, a value that is consistent
with the velocity in the simulation shown in figure
\ref{fig_SimulCL}. The same redshift difference can be 
observed when looking at the surrounding galaxies as can be 
seen form figure \ref{fig_HistoZ} suggesting that MACSJ0416 consists of
two clusters moving against each other in a direction close to the
line of sight and that are close enough so that the X-ray plasma is
already undergoing dynamical friction.

\section{A very flat core region}

\begin{figure}  
   \includegraphics[width=8cm]{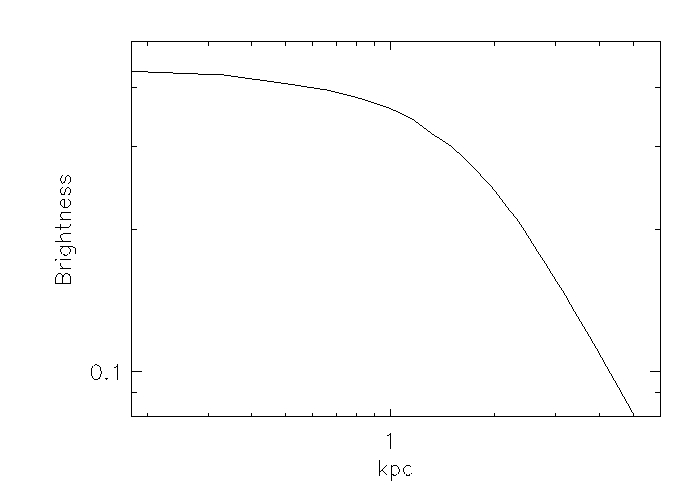} 
   \caption{Brightness profile of the NE cD galaxy showing the break in the 
            Sersic profile at around 1 kpc. The brightness is in arbitrary units.}
    \label{fig_Brightnes}  
\end{figure}

This cluster has an unusual flat profile at around 10-20 arcseconds.
In Z13 it was already shown that the projected mass profile between 10
and 20 arcsec is extremely shallow.  Our model confirms this result as
shown in figure \ref{fig_Profiles} with a break in the slope at a
radius of $\approx 20$ arcsec (about 80 kpc/h). The result is even
more significant if the baryonic component is subtracted from the mass
profile.  In the very central region, and depending on the assumed
model for the fiducial galaxies, the total reconstructed mass
gradually increases again towards the centre.  However, although the
lensing data lacks the sensitivity to constrain the very central
region ($r < 1$ kpc) we should expect an additional flattening of the
profile in the very centre at least for the NE galaxy. This is
suggested by the significant flattening that is also observed in the
light profile of the NE cD galaxy as shown in figure
\ref{fig_Brightnes} (meanwhile the SW cD galaxy does not show this
feature, at least not in a pronounced way).  The flattening in the
galaxy is probably the result of scouring by a massive black hole (or
black hole binary) at the centre of the galaxy
\citep{Postman2012a,Rusli2013,Thomas2014,Lopez2014}.  Stars in radial
trajectories that approach the centre of the galaxy would be ejected
by the black hole(s) towards larger radii, flattening the brightness
profile of the galaxy \citep{Thomas2014}.  The impact parameter,
$r_b$, defined as the radius at which the Sersic profile breaks, is
about $r_b \approx 1$ kpc (see figure \ref{fig_Brightnes}). According
to \cite{Rusli2013}, the radius $r_b$ scales with the mass of the
black hole.  A similar correlation is observed between the black hole
mass and the velocity dispersion of the galaxies
\cite{Gebhardt2000}. For the observed $r_b \approx 1$ kpc, the scaling
laws in \cite{Rusli2013} predict a mass for the super massive black
hole of $M_{BH} \approx 10^10 M_{\odot}$. The existence of the black
hole is supported as well by a very small enhancement of the X-ray
signal at the centre of the NE cD galaxy.

The same black hole that scours the centre of the galaxy of stars
would flatten as well the distribution of dark matter particles
reducing the fraction of radial orbits towards the centre of the
galaxy.  Hence, we should expect also a flattening in the dark matter
distribution in the very central region ($r < 1$ kpc). However, this
mechanism could not explain the flattening at larger radii ($15-100$
kpc/h).  More energetic phenomena, probably involving the past
collision of the two clusters, are required in order to explain the
plateu at $r \approx 10-20$ arcseconds (40-80 kpc/h).  Other
mechanisms play a role in determining the intrinsic shape of the
profile in the central region.  Adiabatic compression of the DM by the
infalling baryons can enhance the densiy profile in the very central
region \cite{Blumenthal1986,Gnedin2004}. On the other hand, this
effect is opposed by the same infalling baryons that can heat up the
core by dynamical friction and can transfer the orbital energy of the
infalling baryons to the DM \cite{ElZant2004}. AGN feedback can
contribute also to the flattening of the central region
\cite{Martizzi2012}. However, all these effects act on scales up to 10-15 kpc and could not explain a flattening 
extending up to 100 kpc.  The dynamics of the merger could have an
impact on larger scales and offer an alternative explanation to the
shape of the profile. However, a flattening of the scale observed in
this cluster is difficult to explain with
simulations. \cite{Ricker2001} studies mergers of clusters for
different mass ratios and impact parameters. Their simulatons include
both DM and baryons. For mass ratios similar to MACSJ0416, they find
that the dark matter profiles are not stronlgly affected after the
collision and the profiles don't show a platau at any radius. 
The same result can be obsereved in our simulations as it is shown in figure
\ref{fig_SimulCL} and also in simulations from our
archive of merging clusters \cite{Molnar2012},
where both clumps conserve their cusps. 
In \cite{Dekel2003}, the authors argue that merging of cuspy
halos inevitably leads to a cusp, thus implying that the two
subclusters in MACSJ0416 must have had their cusps supressed before
the merger by some mechanism.  Simulations of collisions of cored
clusters (with cores of about 100 kpc), suggest that the core can
survive the collision \cite{Ritchie2002}. However, these simulations
do not explain how the core is formed.  Despite the need of
simulations to understand the role of merger dynamics on the profiles
of cluster similar to MACSJ05416, simulations are often difficult to
interpret, specially in the central region of clusters. Different
codes may produce different results depending on their implicit
assumptions. As shown in \cite{Mitchell2009}, smoothed particle
hydrodynamics (or SPH) N-body simulations tend to produce more massive
cores than Eulerian mesh codes although the discrepancy gets reduced
when the refinement level is increased in the mesh code.

Alternatively, if the DM is self-interacting, the probability of interaction increases
with the speed of the DM particles. A cluster collision offers a
unique opportunity to boost the chance of interaction from the
increased relative velocity of the interacting particles.  One
interesting clue can be found in \cite{Rocha2013}. In that work, the
authors simulate large scale structure under the assumption that dark
matter can self-interact (with a higher probability at the cores of
haloes).  Interestingly, for cross-sections $\sigma/m \approx 1
cm^2/g$ they find that in halos of masses $M_{vir} = 2 \times 10^{14}
M_{\odot}$ the profile exhibits a plateu corresponding to core scales
of about 150 kpc.  The plateu extends from the centre up to about 70
kpc and at the centre the density is about an order of magnitde lower
than in the corresponding standard NFW profile.  In our case, if the
impact parameter of the two clusters is about 200 kpc, the chance of
interaction between DM particles should be smaller than in the case of
a head-on collision.  However, it is still important to emphasize how
the characteristics of the cored profiles in \cite{Rocha2013}
strikingly resemble the observed profiles in MACSJ0416 so this
possibility still offers a possible interesting explanation.
Moreover, the fact that this cluster went through a collision would
imply that even a smaller than 1 $cm^2/g$ cross section could be
enough to cause a similar effect since the increase in relative
velocity due to the collision could compensate the decrease in the
cross-section.  A lower than 1 $cm^2/g$ value for $\sigma/m$ is also
prefereed by observations of more relaxed clusters where such a
flatenning of the central region is not observed.

Finally, another probable mechanism that can contribute to disrupting the core are tidal forces. 
In figure \ref{fig_SimulCL}, the DM haloes show an elongation. Since the initial model is symmetric 
the tidal effects must be present at the time we are witnessing this encounter. 
The effect that tidal forces have on stretching the cores of merging clusters will be studied in a subsequent paper. 
 
It is not clear that the extent of these effects would still be
sufficient to explain the observed profiles.  More likely, a
combination of different effects would be the most plausible
explanation. If DM is self-interacting, a small cross section together
with some of the mechanisms described above plus projection effects
could lead to a flattening of the central region in colliding
clusters.  If DM self-interaction plays a significant role, similar
shallow profiles should be observed in other post-collision clusters,
in particular in the morpholically disturbed FF clusters, where the DM
distribution can be inferred directly from gravitational lensing.
Improved observations on this cluster will be available soon from a
multiwavelengths perspective, in the optical with the remaining FF
data (improved photo-z for the lensing systems and potentially new
systems), in X-ray with already planned Chandra observations may
reveal shocks, and a better location of the peaks in the X-ray
emission; in radio with also planned deep VLA observations may reveal
possible evidence of past merging activity and confirm the shock
activity. Together they will throw some light on this interesting
cluster.

\section{Conclusions}
We derive a free-form solution of MACSJ0416 that resembles those obtained with parametric methods. 
To first order, we find that the mass traces the light in this merging and elongated cluster. 
Some offsets beteen the DM and plasma are observed, in particular a very shallow soft dark matter halo surrounding 
each cD galaxy. The intrinsic shape of the soft haloes does not show the intrinsic symmetries assumed in 
parametric methods highlighing the need for free-form methods in these type of clusters. 
Our DM reconstructed distribution shows an elongation in the direction of a peak in the X-ray data suggesting 
that the lensing distortions may be sensitive also to the mass of the plasma. 
The two density profiles associated to each subhalo are almost identical suggesting a 1:1 ratio for this 
merging cluster. The profiles also exhibits a plateau at around 40-100 kpc which could be interpreted as the 
result of dynamical distortions of the subcluster cluster profiles, or projection effects but also 
as possible evidence of self interacting DM with an increased probability of interaction during the collision.
Detailed lensing observations of merging galaxy 
clusters, like those from the FF, together with more detailed simulations of  merging clusters may help clarify 
this situation.

\section{Acknowledgments}  
J.M.D acknowledges support of the consolider project CAD2010-00064 and 
AYA2012-39475-C02-01 funded by the Ministerio de Economia y Competitividad. 
JMD also acknowledges the hospitality of the Department of Physics and Astronomy at UPenn during 
part of this research. 
The scientific results reported in this article are based in part on data
obtained from the Chandra Data Archive.
We would like to thank Harald Ebeling for providing us with the code {\small ASMOOTH} \citep{Ebeling2006} 
that was used to smooth the Chandra data. 


\bsp  
\label{lastpage}

\bibliographystyle{mn2e}
\bibliography{MyBiblio} 

\appendix

\section{Compilation of arc stamps and positions}

\begin{figure*}  
\centerline{ \includegraphics[width= 16cm]{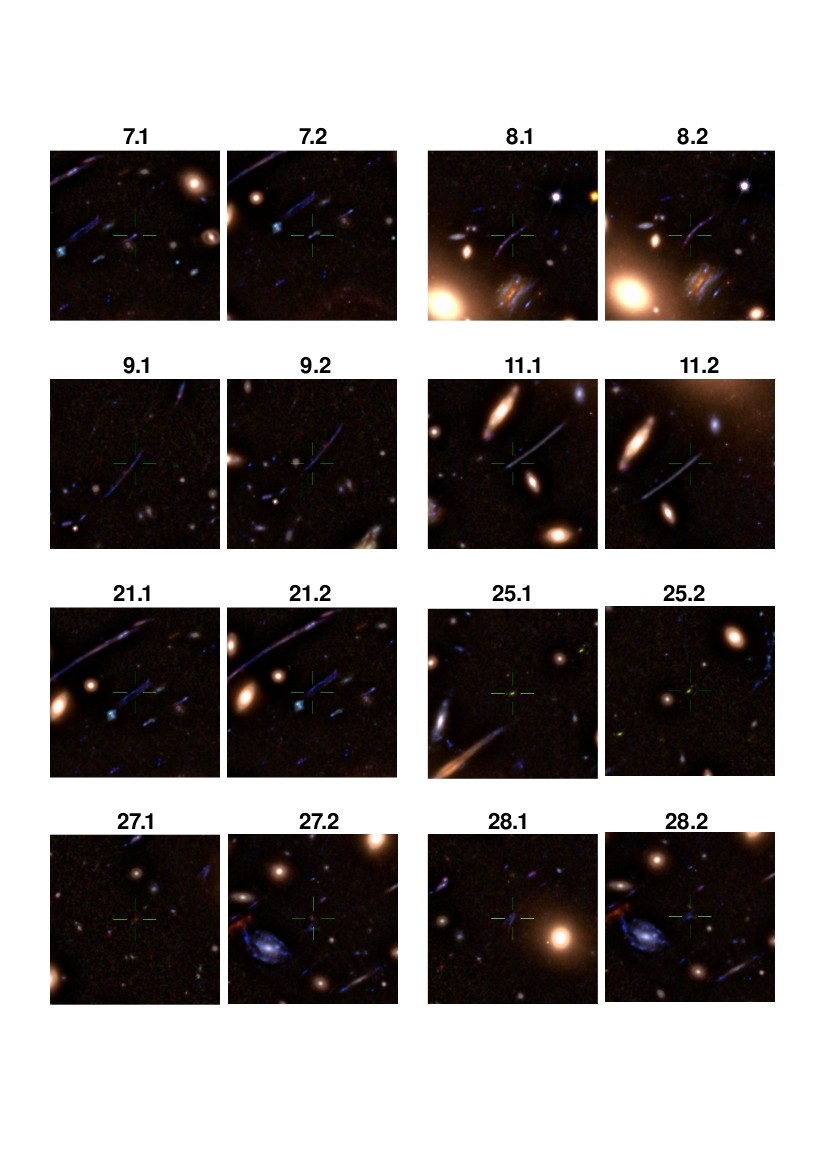}}          
   \caption{Images have been filtered to reduce light glare from member 
            galaxies.}
   \label{fig_StampsI}  
\end{figure*}  

\begin{figure*}  
\centerline{ \includegraphics[width=16cm]{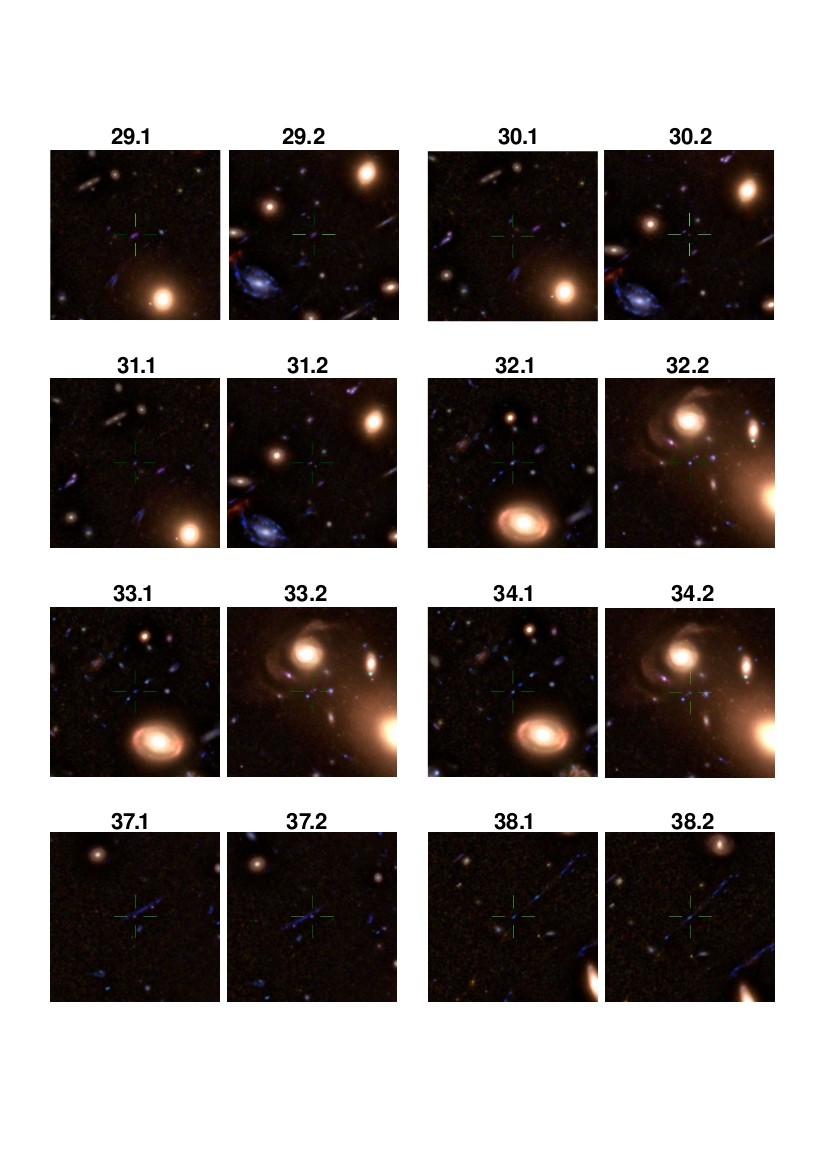}}          
   \caption{Light from member galaxies has been reduced through a high-pass 
            filter.}
   \label{fig_StampsII}  
\end{figure*}  

\begin{figure*}  
\centerline{ \includegraphics[width=16cm]{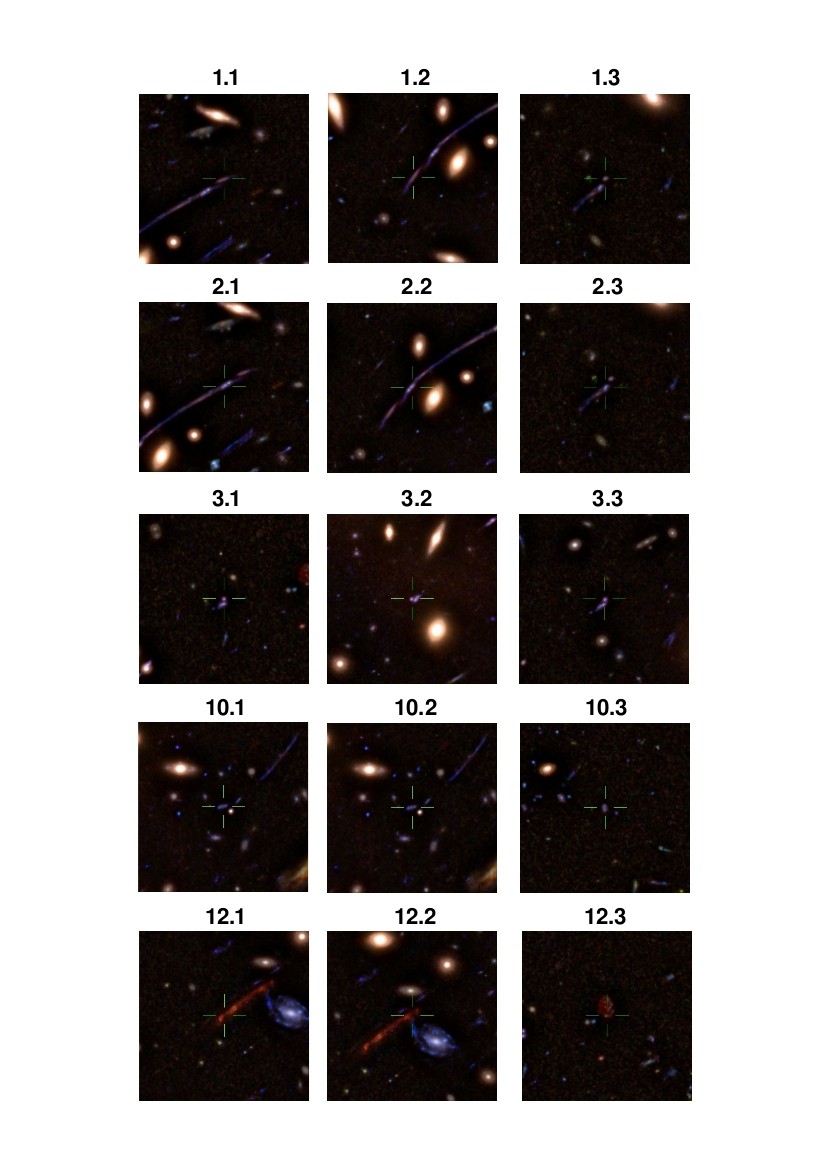}}          
   \caption{Light from member galaxies has been reduced through a high-pass 
            filter.}
   \label{fig_StampsIII}  
\end{figure*}  

\begin{figure*}  
\centerline{ \includegraphics[width=16cm]{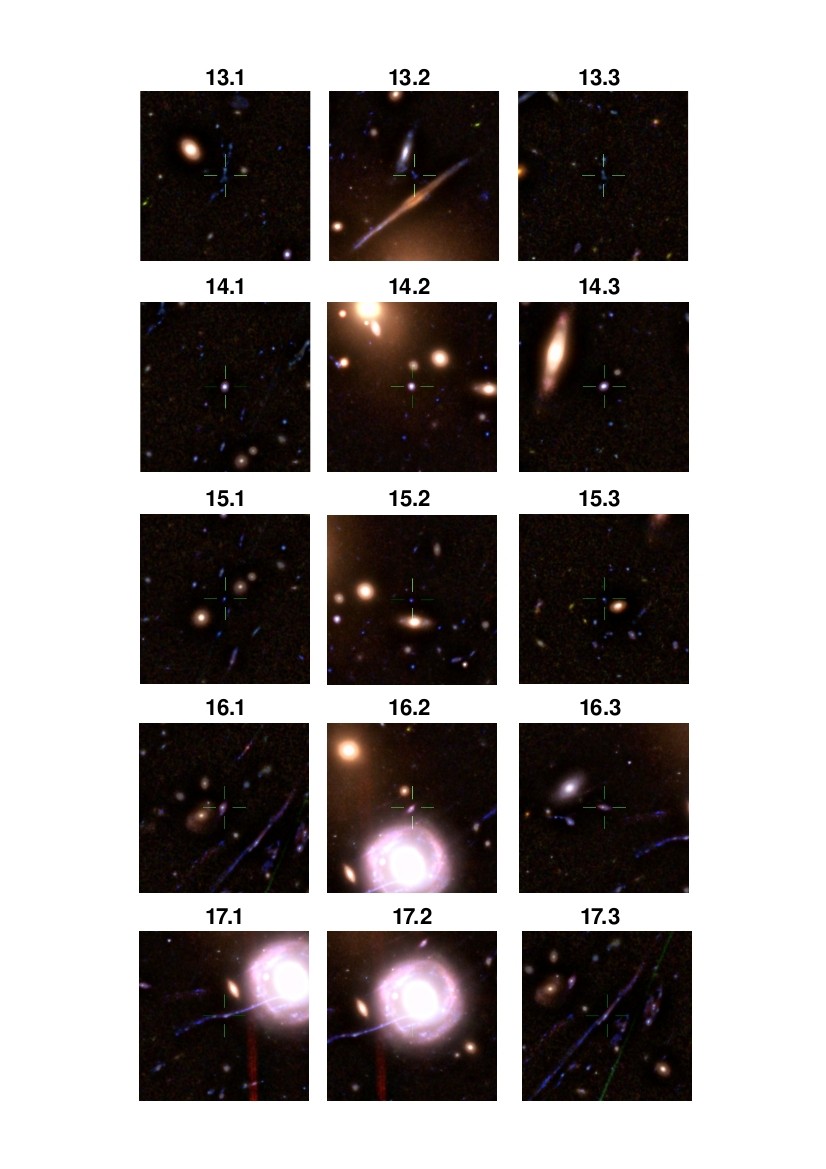}}          
   \caption{Light from member galaxies has been reduced through a high-pass 
            filter.}
   \label{fig_StampsIV}  
\end{figure*}  

\begin{figure*}  
\centerline{ \includegraphics[width=16cm]{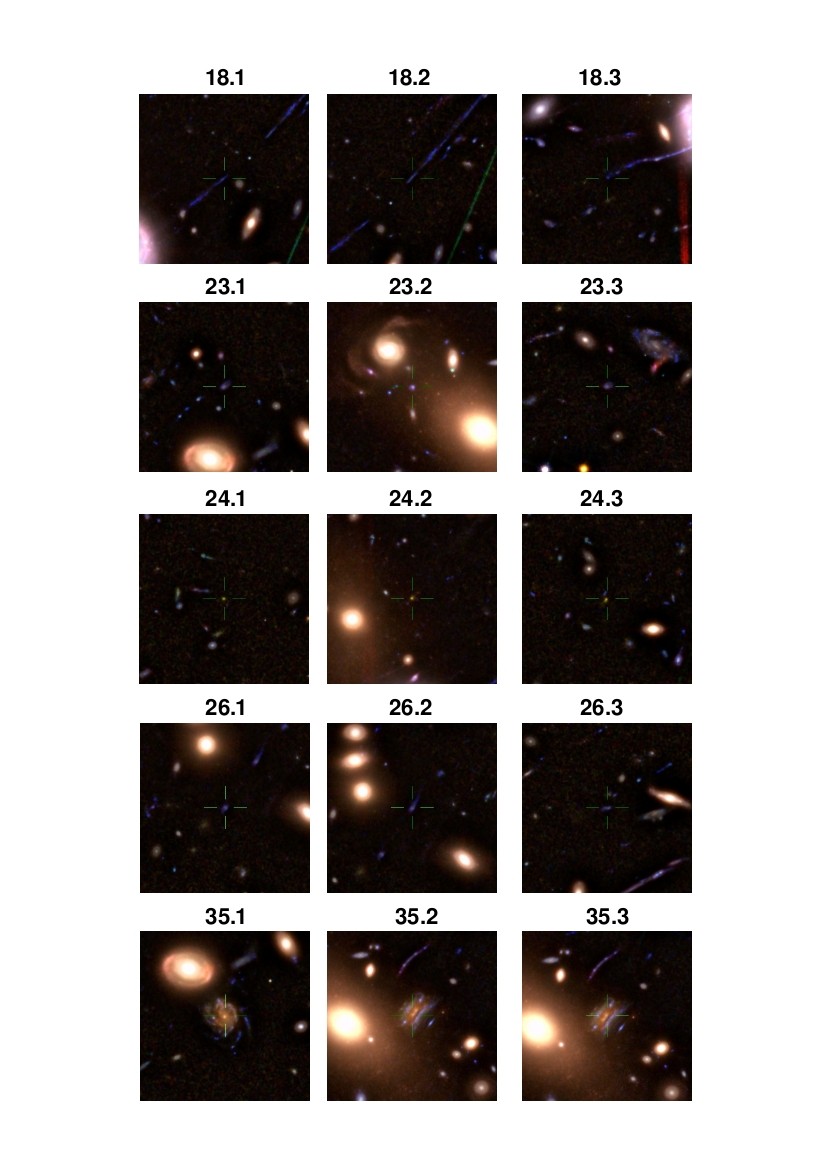}}          
   \caption{Light from member galaxies has been reduced through a high-pass 
            filter.}
   \label{fig_StampsV}  
\end{figure*}  

\begin{figure*}  
\centerline{ \includegraphics[width=16cm]{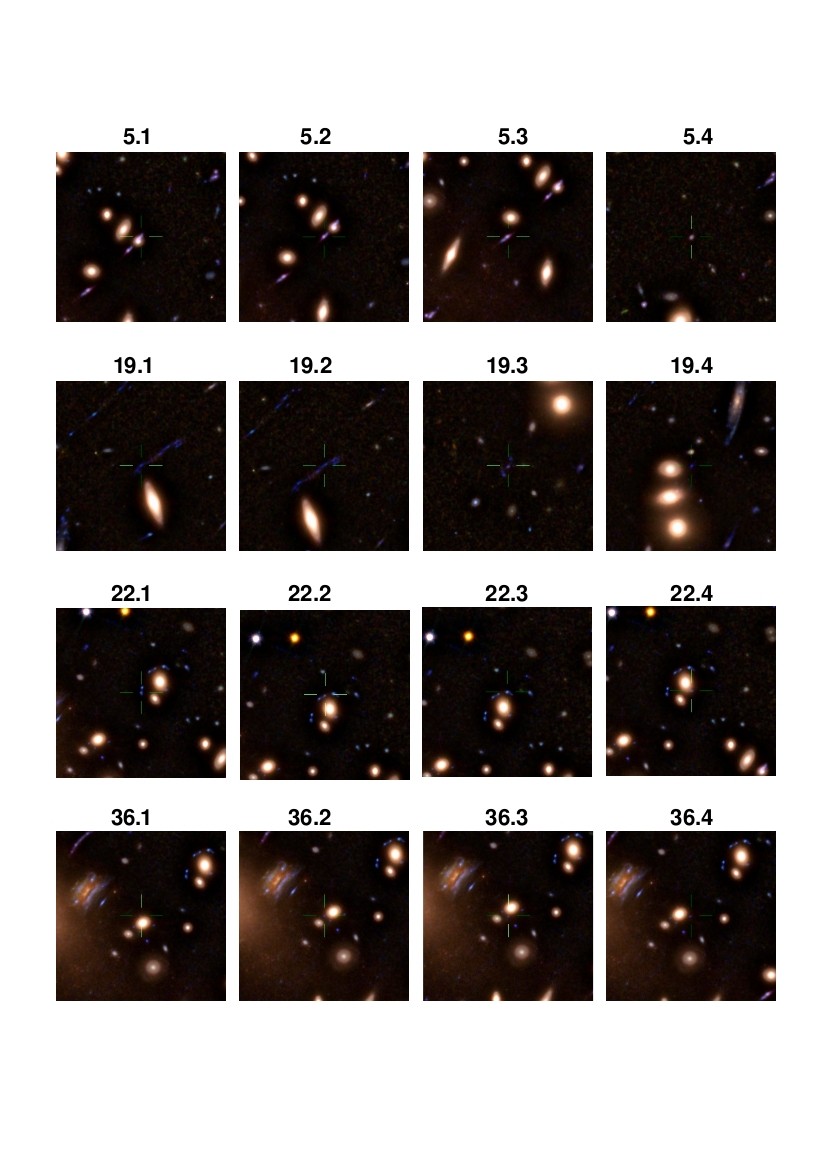}}          
   \caption{Light from member galaxies has been reduced through a high-pass 
            filter.}
   \label{fig_StampsVI}  
\end{figure*}  

\begin{figure*}  
\centerline{ \includegraphics[width=16cm]{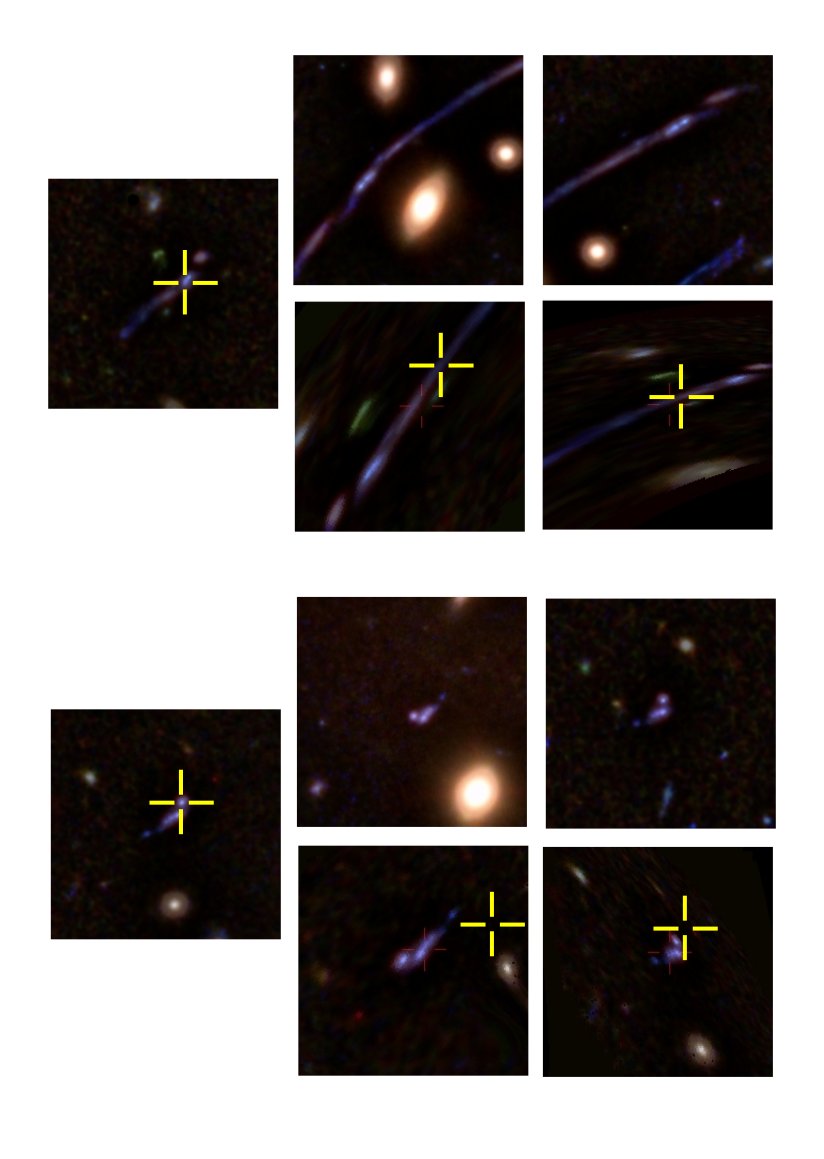}}          
   \caption{Top group, relensed system 1 (and 2). The image on the left is the one used to delense the galaxy 
            into the source plane. On the right we show the observed (top) and predicted (bottom) counterimages.
            Bottom group, like the top group but for the relensed system 3 (and 4).
            The crosses mark the observed positions. We use the model in case ii) for the predicted images.  
            In all cases the field of view is 7.2 arcseconds.}
   \label{fig_Relensed_1to4}  
\end{figure*}

    \begin{table*}
    \begin{minipage}{115mm}                                               
    \caption{Full strong lensing data set. The first column shows system ID following the original notation 
             of Z13.
             The second and third columns show the coordinates of each arclet. 
             Column 4 includes the redshifts used in our study (taken from Z13 when appropriate).
             Some of these redshifts are photometric and some are based on color and/or predicted by the lens model. 
	     The last column contains additional useful information. In this column, gz indicates that the redshift 
             has been derived by the lens model.
             System 4 and system 3 are almost overlapping in position and could be in fact the same system 
             but are treated as two different systems following Z13. 
             Candidates marked with (*) correspond to 
             new arclet candidates identified in this work. Arclet 12.3 is found at the position predicted by 
             Z13. Due to the possibility of confusion by multiple faint arclets with similar colors 
             and falling into the same region, we have not used the following arclets from Z13 7.3, 9.3
             10.3,11.3, 12.3(updated), 15.3(updated), 21.3. Systems 6 and 20 from Z13 
             have been discarded completely due to bad color agreement after adding the new IR data.}
 \label{tab1}
 \begin{tabular}{ccccc}   

  ID     &  RAJ2000(h:m:s)  & DECJ2000(d:m:s)  &    z    & Notes  \\

  1.1    &   04:16:09.780   &   -24:03:41.73   &  1.896  &    spect z \\
  1.2    &   04:16:10.435   &   -24:03:48.75   &  1.896  & \\   
  1.3    &   04:16:11.365   &   -24:04:07.21   &  1.896  & \\   
  2.1    &   04:16:09.884   &   -24:03:42.77   &  1.8    & \\   
  2.2    &   04:16:10.321   &   -24:03:46.93   &  1.8    & \\  
  2.3    &   04:16:11.394   &   -24:04:07.86   &  1.8    & \\  
  3.1    &   04:16:07.388   &   -24:04:01.62   &  2.25   & \\  
  3.2    &   04:16:08.461   &   -24:04:15.53   &  2.25   & \\
  3.3    &   04:16:10.036   &   -24:04:32.56   &  2.25   & \\  
  4.1    &   04:16:07.398   &   -24:04:02.01   &  2.25   &   Same as 3.1? \\
  4.2    &   04:16:08.437   &   -24:04:15.53   &  2.25   &   Same as 3.2? \\
  4.3    &   04:16:10.051   &   -24:04:33.08   &  2.25   &   Same as 3.3? \\
  5.1    &   04:16:07.773   &   -24:04:06.24   &  2.3    & \\    
  5.2    &   04:16:07.839   &   -24:04:07.21   &  2.3    & \\  
  5.3    &   04:16:08.043   &   -24:04:10.01   &  2.3    & \\  
  5.4    &   04:16:10.454   &   -24:04:37.05   &  2.3    & \\
  7.1    &   04:16:09.552   &   -24:03:47.13   &  2.0    & \\  
  7.2    &   04:16:09.752   &   -24:03:48.82   &  2.0    & \\  
  8.1    &   04:16:08.783   &   -24:03:58.05   &  2.4    & \\
  8.2    &   04:16:08.840   &   -24:03:58.83   &  2.4    & \\  
  9.1    &   04:16:06.486   &   -24:04:42.90   &  2.3    & \\  
  9.2    &   04:16:06.605   &   -24:04:44.78   &  2.3    & \\  
  10.1   &   04:16:06.244   &   -24:04:37.76   &  2.3    & \\
  10.2   &   04:16:06.833   &   -24:04:47.12   &  2.3    & \\  
  10.3   &   04:16:08.807   &   -24:05:01.94   &  2.3    & \\ 
  11.1   &   04:16:09.410   &   -24:04:13.32   &  0.8    & \\
  11.2   &   04:16:09.196   &   -24:04:11.11   &  0.8    & \\
  12.1   &   04:16:09.230   &   -24:04:25.74   &  2.5    & \\ 
  12.2   &   04:16:09.011   &   -24:04:23.72   &  2.5    & \\ 
  12.3   &   04:16:06.956   &   -24:04:00.57   &  2.5    &        New(p12.3)  \\
  13.1   &   04:16:06.619   &   -24:04:22.03   &  3.0    & \\
  13.2   &   04:16:07.711   &   -24:04:30.61   &  3.0    & \\  
  13.3   &   04:16:09.681   &   -24:04:53.56   &  3.0    & \\ 
  14.1   &   04:16:06.296   &   -24:04:27.62   &  1.7    & \\ 
  14.2   &   04:16:07.450   &   -24:04:44.26   &  1.7    & \\ 
  14.3   &   04:16:08.598   &   -24:04:52.78   &  1.7    & \\ 
  15.1   &   04:16:06.292   &   -24:04:33.67   &  2.0    & \\  
  15.2   &   04:16:07.065   &   -24:04:42.90   &  2.0    & \\  
  15.3   &   04:16:09.175   &   -24:04:58.71   &  2.0    &        New(*)  \\
  16.1   &   04:16:05.774   &   -24:04:51.22   &  2.0    & \\ 
  16.2   &   04:16:06.799   &   -24:05:04.35   &  2.0    & \\     
  16.3   &   04:16:07.583   &   -24:05:08.77   &  2.0    & \\ 
  17.1   &   04:16:07.170   &   -24:05:10.91   &  2.5    & \\
  17.2   &   04:16:06.866   &   -24:05:09.55   &  2.5    & \\ 
  17.3   &   04:16:05.599   &   -24:04:53.69   &  2.5    & \\ 
  18.1   &   04:16:06.258   &   -24:05:03.24   &  2.8    & \\ 
  18.2   &   04:16:06.016   &   -24:05:00.06   &  2.8    & \\ 
  18.3   &   04:16:07.416   &   -24:05:12.28   &  2.8    & \\      
  19.1   &   04:16:10.909   &   -24:03:41.08   &  2.7    & \\     
  19.2   &   04:16:10.777   &   -24:03:39.85   &  2.7    & \\ 
  19.3   &   04:16:11.925   &   -24:04:00.91   &  2.7    & \\ 
  19.4   &   04:16:11.580   &   -24:03:52.23   &  2.7    &       New(*)  \\

 \end{tabular}
 \end{minipage}
\end{table*}

\setcounter{table}{0}
    \begin{table*}
    \begin{minipage}{115mm}                                               
    \caption{cont.}

 \begin{tabular}{ccccc}    

  ID     &  RAJ2000(h:m:s)  & DECJ2000(d:m:s)  &    z    & Notes  \\

  21.1   &   04:16:09.813   &   -24:03:46.67   &  2.6    & \\ 
  21.2   &   04:16:09.865   &   -24:03:47.32   &  2.6    & \\ 
  22.1   &   04:16:08.278   &   -24:04:01.07   &  1.8    & \\ 
  22.2   &   04:16:08.204   &   -24:03:59.28   &  1.8    & \\    
  22.3   &   04:16:08.162   &   -24:03:59.22   &  1.8    & \\ 
  22.4   &   04:16:08.152   &   -24:04:00.87   &  1.8    &       New(*) \\
  23.1   &   04:16:10.691   &   -24:04:19.56   &  2.0    & \\ 
  23.2   &   04:16:09.505   &   -24:03:59.87   &  2.0    & \\ 
  23.3   &   04:16:08.242   &   -24:03:49.47   &  2.0    & \\ 
  24.1   &   04:16:08.413   &   -24:05:07.85   &  5.0    &       New   \\
  24.2   &   04:16:06.820   &   -24:04:58.83   &  5.0    &       New   \\
  24.3   &   04:16:05.517   &   -24:04:38.16   &  5.0    &       New   \\
  25.1   &   04:16:07.385   &   -24:04:27.03   &  4.0    &       New   \\
  25.2   &   04:16:07.028   &   -24:04:24.00   &  4.0    &       New   \\
  26.1   &   04:16:11.551   &   -24:04:01.05   &  2.0    &       New   \\
  26.2   &   04:16:11.394   &   -24:03:57.75   &  2.0    &       New   \\
  26.3   &   04:16:10.136   &   -24:03:37.95   &  2.0    &       New   \\
  27.1   &   04:16:10.056   &   -24:04:38.79   &  4.0    &       New   \\
  27.2   &   04:16:08.658   &   -24:04:23.49   &  4.0    &       New   \\
  28.1   &   04:16:09.683   &   -24:04:35.37   &  2.0    &       New   \\
  28.2   &   04:16:08.662   &   -24:04:24.15   &  2.0    &       New   \\
  29.1   &   04:16:09.578   &   -24:04:32.19   &  2.0    &       New   \\
  29.2   &   04:16:08.605   &   -24:04:22.11   &  2.0    &       New   \\
  30.1   &   04:16:09.705   &   -24:04:32.79   &  2.0    &       New   \\
  30.2   &   04:16:08.632   &   -24:04:20.88   &  2.0    &       New   \\
  31.1   &   04:16:09.718   &   -24:04:31.71   &  2.0    &       New   \\
  31.2   &   04:16:08.651   &   -24:04:20.64   &  2.0    &       New   \\
  32.1   &   04:16:10.824   &   -24:04:20.43   &  2.0    &       New   \\
  32.2   &   04:16:09.620   &   -24:04:00.27   &  2.0    &       New   \\
  33.1   &   04:16:10.886   &   -24:04:21.15   &  2.0    &       New   \\
  33.2   &   04:16:09.598   &   -24:03:59.97   &  2.0    &       New   \\
  34.1   &   04:16:10.921   &   -24:04:21.63   &  2.0    &       New   \\
  34.2   &   04:16:09.582   &   -24:03:59.82   &  2.0    &       New   \\
  35.1   &   04:16:10.575   &   -24:04:28.02   &  1.65   &       New gz   \\
  35.2   &   04:16:08.803   &   -24:04:02.43   &  1.65   &       New gz  \\
  35.3   &   04:16:08.780   &   -24:04:02.06   &  1.65   &       New gz  \\
  36.1   &   04:16:08.509   &   -24:04:03.86   &  2.0    &       New   \\
  36.2   &   04:16:08.547   &   -24:04:04.64   &  2.0    &       New   \\
  36.3   &   04:16:08.514   &   -24:04:04.95   &  2.0    &       New   \\
  36.4   &   04:16:08.441   &   -24:04:04.49   &  2.0    &       New   \\
  37.1   &   04:16:07.931   &   -24:04:54.42   &  2.0    &       New   \\
  37.2   &   04:16:07.841   &   -24:04:53.73   &  2.0    &       New   \\
  38.1   &   04:16:11.271   &   -24:03:38.85   &  2.0    &       New   \\
  38.2   &   04:16:11.148   &   -24:03:37.41   &  2.0    &       New   \\
 \end{tabular}
 \end{minipage}
\end{table*}

    \begin{table*}
    \begin{minipage}{115mm}                                               
    \caption{Arclets from Z13 that are not used in this work. Systems 6 and 20 where removed completely.}
 \label{tab1}
 \begin{tabular}{ccccc}   

  ID     &  RAJ2000(h:m:s)  & DECJ2000(d:m:s)  & Notes  \\

  7.3    &  04:16:11.308   &  -24:04:15.99     & Many candidates nearby  \\
  9.3    &  04:16:09.149   &  -24:05:01.23     & Many candidates nearby. Far from prediction  \\
  10.3   &  04:16:09.818   &  -24:04:58.69     & The alternative 10.3 works better \\
  11.3   &  04:16:08.214   &  -24:03:57.66     & No predicted 3rd counterimage  \\
  12.3   &  04:16:06.989   &  -24:04:03.57     & Red galaxy nearby fits position better (and color) \\
  15.3   &  04:16:08.560   &  -24:04:55.38     & Far from model predcition \\
  21.3   &  04:16:11.047   &  -24:04:07.73     & Far from model prediction \\

 \end{tabular}
 \end{minipage}
\end{table*}

\end{document}